\documentclass{aamod}

\usepackage{natbib}
\bibpunct{(}{)}{;}{a}{}{,}
\usepackage{graphicx}
\usepackage{txfonts}
\usepackage{lscape}

\begin{document}

\title{Magnetic structure of an activated filament in a flaring active region}

\author{C. Sasso\inst{1,2} \and A. Lagg\inst{1} \and S. K. Solanki\inst{1,3}}

\offprints{C. Sasso, \email{csasso@oacn.inaf.it}}

\institute{Max-Planck-Institut f\"ur Sonnensystemforschung,
Max-Planck-Str. 2, D-37191 Katlenburg-Lindau, Germany \and INAF-Osservatorio
Astronomico di Capodimonte, Salita Moiariello 16, I-80131 Napoli,
Italy \and School of Space Research, Kyung Hee University, Yongin, Gyeonggi 
446-701, Republic of Korea}

\date{Received / Accepted}

\abstract{}{While the magnetic field in quiescent prominences has been widely 
investigated, less is known about the field in activated prominences. We 
introduce observational results on the magnetic field structure of an 
activated filament in a flaring active region. We study, in particular, its 
magnetic structure and line-of-sight flows during its early activated phase,
shortly before it displays signs of rotation.}{We invert the Stokes profiles
of the chromospheric \ion{He}{i} 10830~{\AA} triplet and the photospheric
\ion{Si}{i} 10827~{\AA} line observed in this filament by the VTT on
Tenerife. Using these inversion results we present and interpret the first
maps of velocity and magnetic field obtained in an activated filament, both in 
the photosphere and the chromosphere.}{Up to 5 different magnetic components
are found in the chromospheric layers of the filament, while outside the
filament a single component is sufficient to reproduce the observations. 
Magnetic components displaying an upflow are preferentially located towards
the centre of the filament, while the downflows are concentrated along its
periphery. Also, the upflowing gas is associated with an opposite-polarity
magnetic configuration with respect to the photosphere, while the downflowing
gas is associated with a same-polarity configuration.}{The activated filament
has a rather complex structure. Nonetheless, it is compatible with a flux 
rope, although with a distorted one, in the \textit{normal} configuration. The 
observations are best explained by a rising flux rope in which a part of the
filament material is still stably stored (upflowing material, rising with the
field), while a part is no longer stably stored and flows down along the field 
lines.}


\maketitle

\section{Introduction}

Filaments are relatively dense and cold objects, embedded in the surrounding 
much hotter and thinner corona \citep{thb}. When seen projected against 
the solar disc, filaments show up in absorption in H$\alpha$, appearing as 
dark, elongated features against the bright disc. They are the on-disc 
counterparts of prominences that appear as bright features against the dark 
background above the solar limb. Filaments are always located above the 
neutral lines that separate regions of opposite magnetic polarity in the
photospheric field \citep{babcock} and beneath the coronal arcades connecting
these opposite polarity regions. 

Although prominences have been studied for decades, there are still many open 
questions regarding their formation, structure and stability. Magnetic fields 
are known to play a key role in supporting the prominence material against 
gravity and to act as a thermal shield for the cool plasma against the 
million-degree coronal environment \citep{kippenhahn,thb}. Different 
theoretical models have been developed to explain how the cold prominence 
plasma can be supported by magnetic field in the solar corona, many of which
propose magnetic configurations with dipped field lines 
\citep{kippenhahn,wu,antiochos2,choe,vanhoven}. The first authors to 
propose and develop a model for the equilibrium and stability of prominence 
plasma in a magnetic configuration with dipped field lines were 
\citet{kippenhahn}. They assumed a main field configuration in which the 
polarity of the magnetic field lines threading the prominence material is the 
same as the polarity of the underlying photospheric field. In this case the 
prominence has a \textit{normal polarity field}. Later, \citet{kuth} proposed 
a different field configuration which was developed and detailed by 
\citet{kuperus}. A closed-loop magnetic field supports the filament material 
and shields it from the hot coronal plasma. In this model the magnetic field 
lines threading the prominence material are in the opposite direction than the 
photospheric magnetic field. In this case the prominence has an 
\textit{inverse polarity field}. 

Both models are two--dimensional, with the magnetic field confined to a 
vertical plane with respect to the solar surface. Different investigations
\citep{hyder,rust,leroy2,leroy} revealed that prominences possess a strong 
magnetic field along the axial direction. These led to the development of 
three--dimensional models: helical flux ropes that lie horizontally above the
polarity inversion line (PIL). The flux rope naturally presents dips (troughs 
of the helical windings) where plasma can accumulate in a stable manner. The 
prominence material sits in the lower part of the flux rope in these troughs. 

The idea of quiescent prominences supported by helical flux ropes that lie 
horizontally above the photospheric PIL, with the main axis roughly 
parallel to the PIL, has been discussed and developed later by, for example, 
\citet{ballegooijen,priesta,priestb}. Research has also moved into the
detailed description of the magnetic topology of the flux rope. A large
twisted magnetic flux rope is indeed an appealing prominence model because its
helical field lines provide support for the mass of the prominence and isolate
the cold prominence plasma from the hot corona. A coronal flux rope can be
defined as a magnetic structure that contains field lines that twist about
each other by more than one winding between the two ends anchored to the
photosphere \citep{priesta,low1}. Most of the present models are based on flux
tubes that possess some degree of twist
\citep{ballegooijen,priesta,priestb,antiochos2,amari}. Such models comprise
also a coronal arcade that overlies the helical field. Depending on the class
of models (normal or inverse polarity) the coronal magnetic field around 
prominences has different characteristics and this likely plays an 
important role in the equilibrium and stability of the prominence field. 

First measurements of the magnetic field strength in prominences found typical
values of 10-20~G and a magnetic field vector lying almost parallel to the 
axis of the main body of the prominence, forming with it an angle of 
$\sim20^\circ$ in the horizontal plane \citep[see][for a review]{leroy}. The 
observations analysed were obtained in the \ion{He}{i} D$_3$ line at 
5876~{\AA} seen in emission in quiescent prominences. Practically, the 
measurements have indicated the predominance of horizontal field lines 
threading the prominence whose principal component is directed along the long 
axis of the prominence.

In recent years, \citet{casini} presented the first map of the magnetic field 
in a prominence. They retrieved the magnetic field vector 
values by inverting spectropolarimetric data obtained in the \ion{He}{i} D$_3$ 
line seen in emission. The average magnetic field is mostly horizontal and 
varies between 10 and 20~G. These results confirm previously measured values. 
However, the maps also show that the field can be significantly stronger than 
average, up to 80~G, in clearly organised plasma structures of the
prominence. They also found a slow turn of the field vector away from the 
prominence axis as the edge of the prominence is approached. Using instead the
\ion{He}{i} triplet at 10830~{\AA}, \citet{Trujillo_nature} found horizontal 
fields with a strength of 20~G in a filament located at the very centre of the 
solar disc.
 
The analysis of polarimetric data has suggested that most quiescent 
prominences have inverse polarity. Only a minority of prominences, generally 
low-lying, shows a normal configuration \citep{bommier}. The measurements also 
show a tendency for the field strength and the alignment of the magnetic 
vector with the prominence axis to increase with height inside prominences 
observed at the limb. These two properties have been explained in terms of the 
magnetic field of a prominence embedded in a flux rope in the inverse 
configuration. These models have been successful in explaining also other 
observational constraints on prominences, including the fact that when a 
prominence erupts it sometimes looks like a twisted tube. 

While the magnetic field in quiescent prominences has been widely 
investigated, few investigations have dealt with the field in active 
prominences or filaments. Thus, \citet{kuckein} and \citet{xu} studied the 
vector magnetic field of active region filaments by analysing
spectropolarimetric data in the \ion{He}{i} 10830~{\AA} lines, finding the
highest field strengths measured in filaments so far, around 600-700~G
\citep[cf.][]{guo}, although parts of the filament may be associated with
weaker fields. \citet{xu} and \citet{kuckein1,kuckein2} performed a
multiheight study of the vector magnetic field in active region filaments,
through spectropolarimetric observations in the chromospheric \ion{He}{i}
10830~{\AA} multiplet and the photospheric \ion{Si}{i} 10827~{\AA} line. The
inferred vector magnetic fields of the filament suggest a flux rope topology,
with the field pointing mainly along the filament. These observations also
indicate that the filament is divided into two parts, one that lies in the
upper chromosphere and another one that stays trapped in the lower
chromosphere/upper photosphere. The complex structure of active region
filaments is suggested also by the emergence of a flux rope below an already
existing filament \citep{okamoto}. Measurements of the magnetic field in
activated or erupting filaments have to our knowledge not been carried out so
far. Such measurements are potentially important for constraining models of
erupting filaments \citep[e.g.,][]{gibson,roussev,suvanba}.

In this paper we introduce observational results on the magnetic field 
measurements in an active region filament that was activated by a flare. We 
provide evidence for the existence of a coronal flux rope in this filament,
albeit a strongly distorted one. Moreover, we are able to support a normal 
polarity configuration of the flux rope structure. The results presented here
build on our previous paper, \citet[hereafter referred to as Paper I]{sasso1}, 
in which we described the inversion of the often broad and very complex Stokes 
profiles of the \ion{He}{i} 10830~{\AA} filament, observed in the studied 
activated filament.

The paper is structured as follows: in Sec.~\ref{sec:obs}, we briefly 
summarise the observations and the results obtained by inverting the Stokes 
profiles. In Sec.~\ref{sec:results}, the obtained maps of the magnetic and 
velocity fields in the filament are presented and critically analysed. 
Finally, in the discussion and conclusions (Sec.~\ref{sec:conlusions}), the 
information obtained from the inversions of the spectropolarimetric data are 
interpreted in terms of filament models.

\section{Observations and data analysis}\label{sec:obs}

On 2005 May 18 we observed the active region NOAA 10763, located at $24^\circ$ 
W, $14^\circ$ S on the solar disk, which corresponds to a cosine of the 
heliocentric angle of $\mu=0.9$. Some minutes before the spectrograph started 
scanning the active region a flare of GOES class C2.0 erupted 
in the western part of the active region. A filament, present in the active 
region close to and partly overlying the flare ribbons, was activated by the 
flare. 

\begin{figure*}
\centering
\includegraphics[width=3.3cm,clip=true]{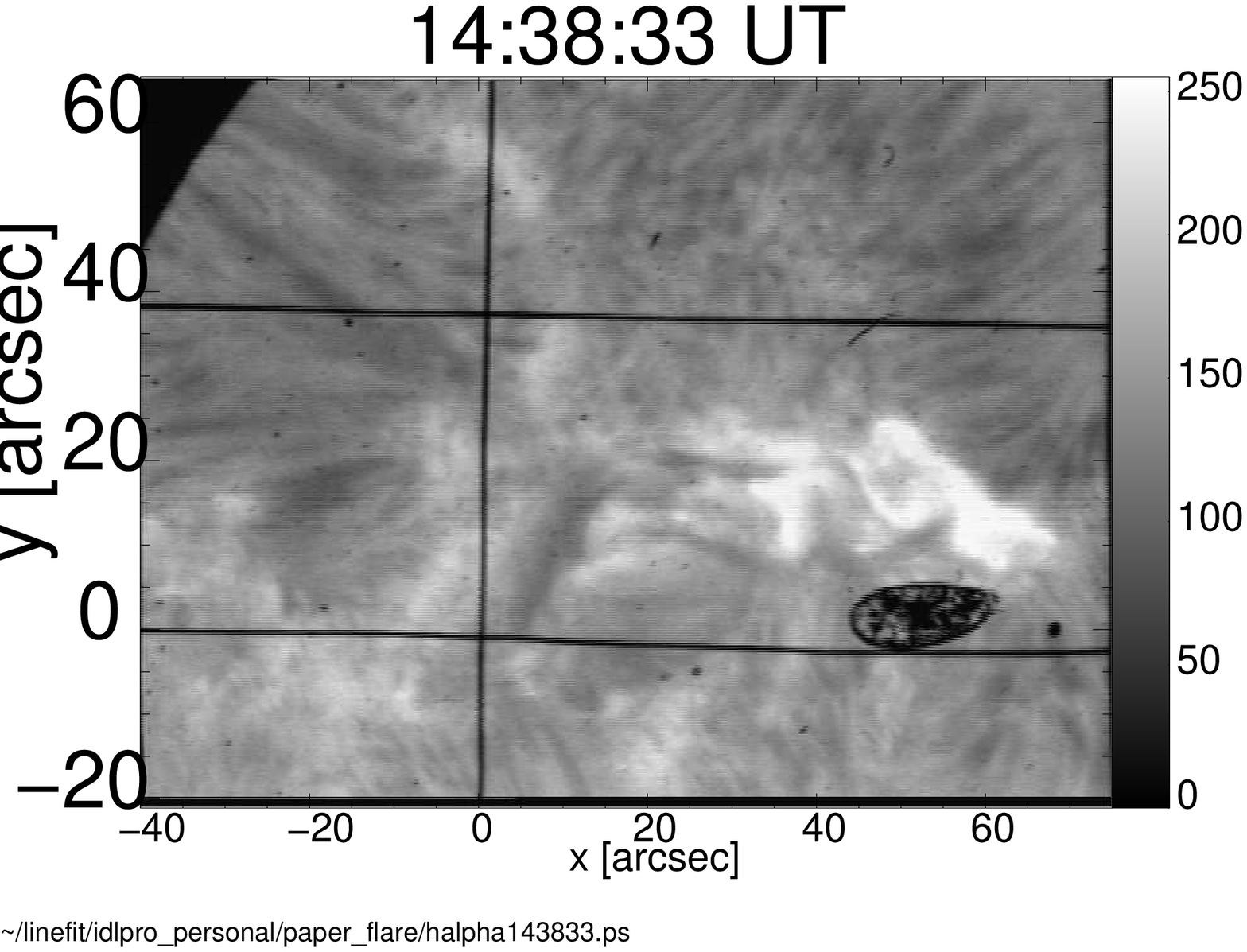}
\includegraphics[width=3.3cm,clip=true]{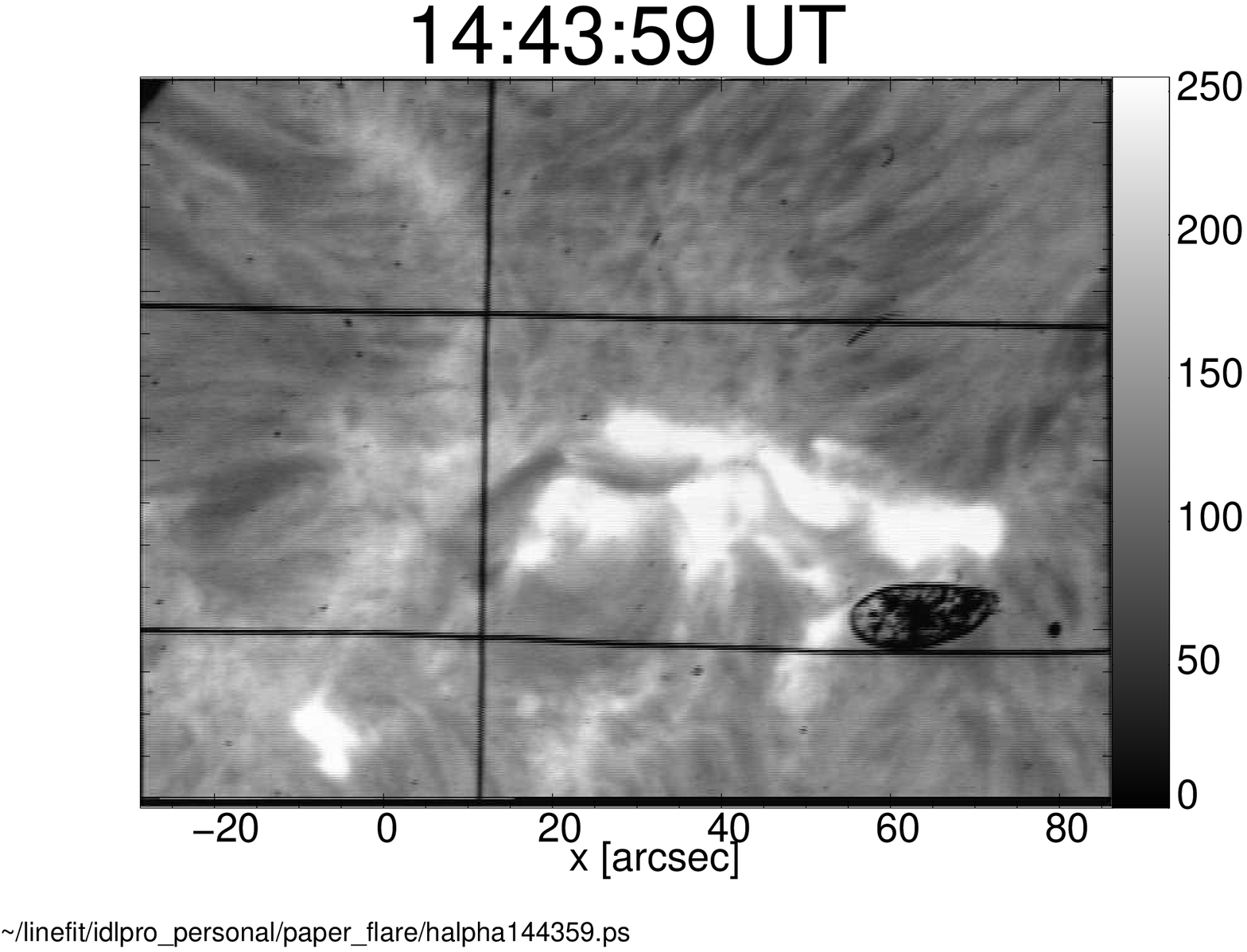}
\includegraphics[width=3.3cm,clip=true]{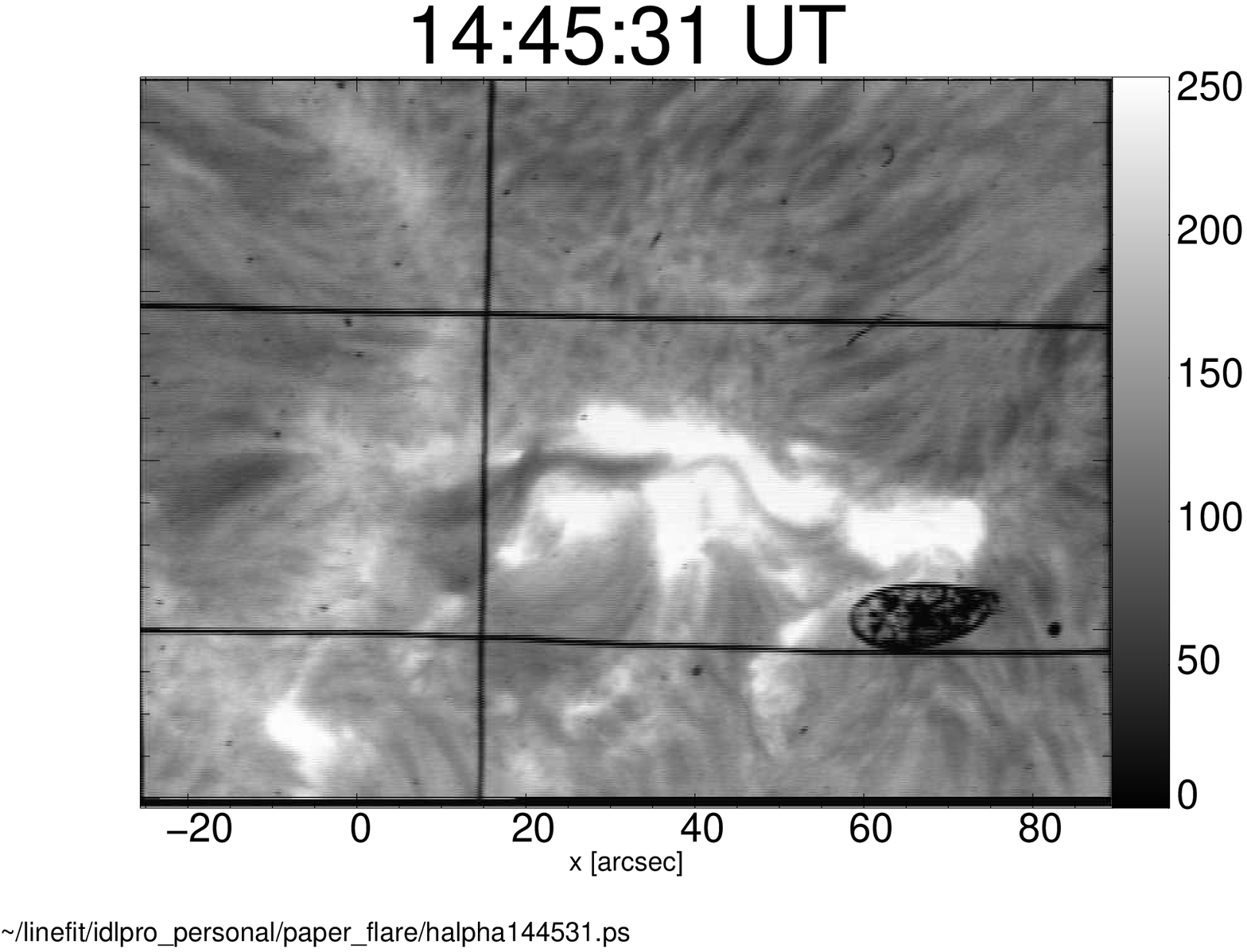}
\includegraphics[width=3.3cm,clip=true]{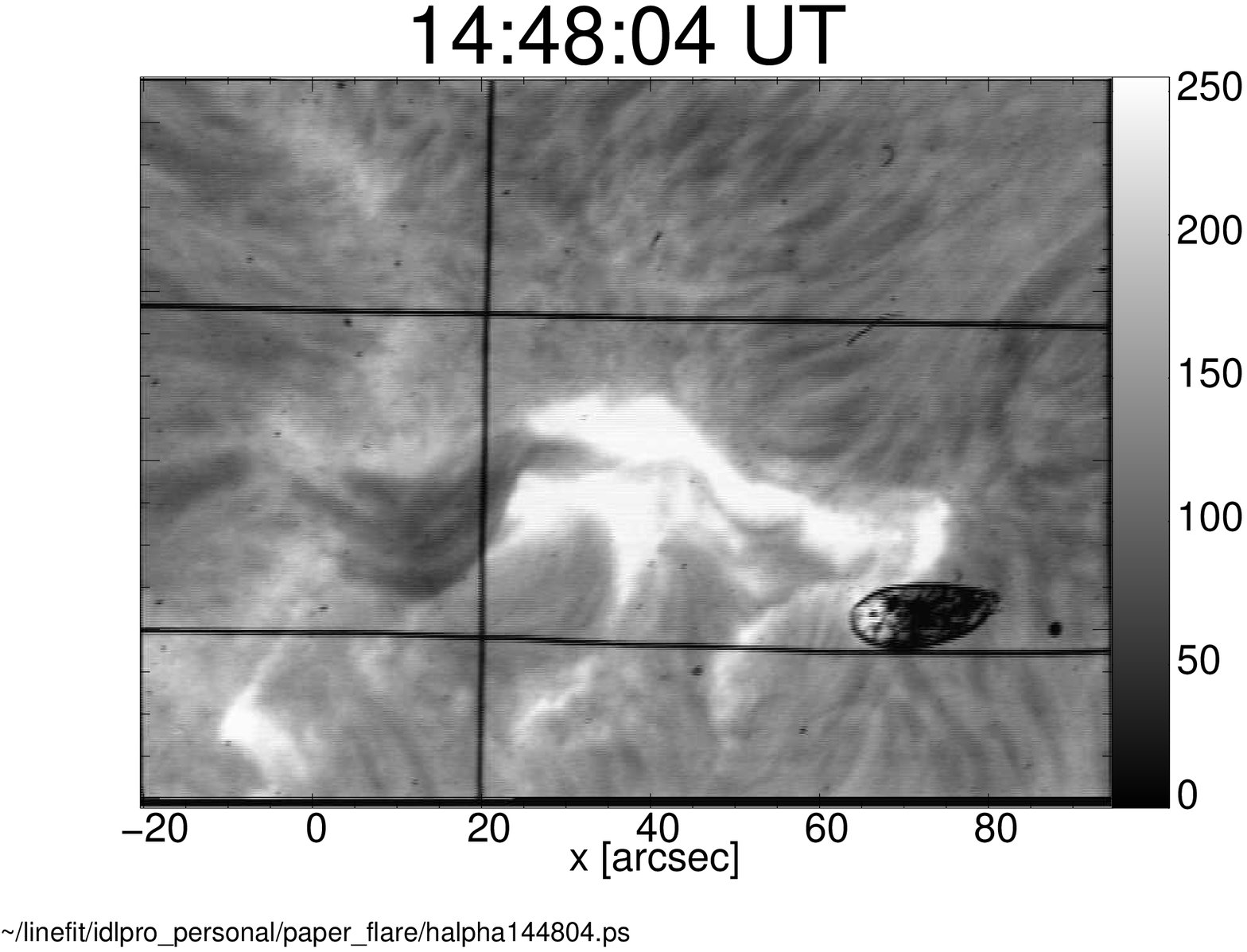}
\hspace{3cm}
\includegraphics[width=3.3cm,clip=true]{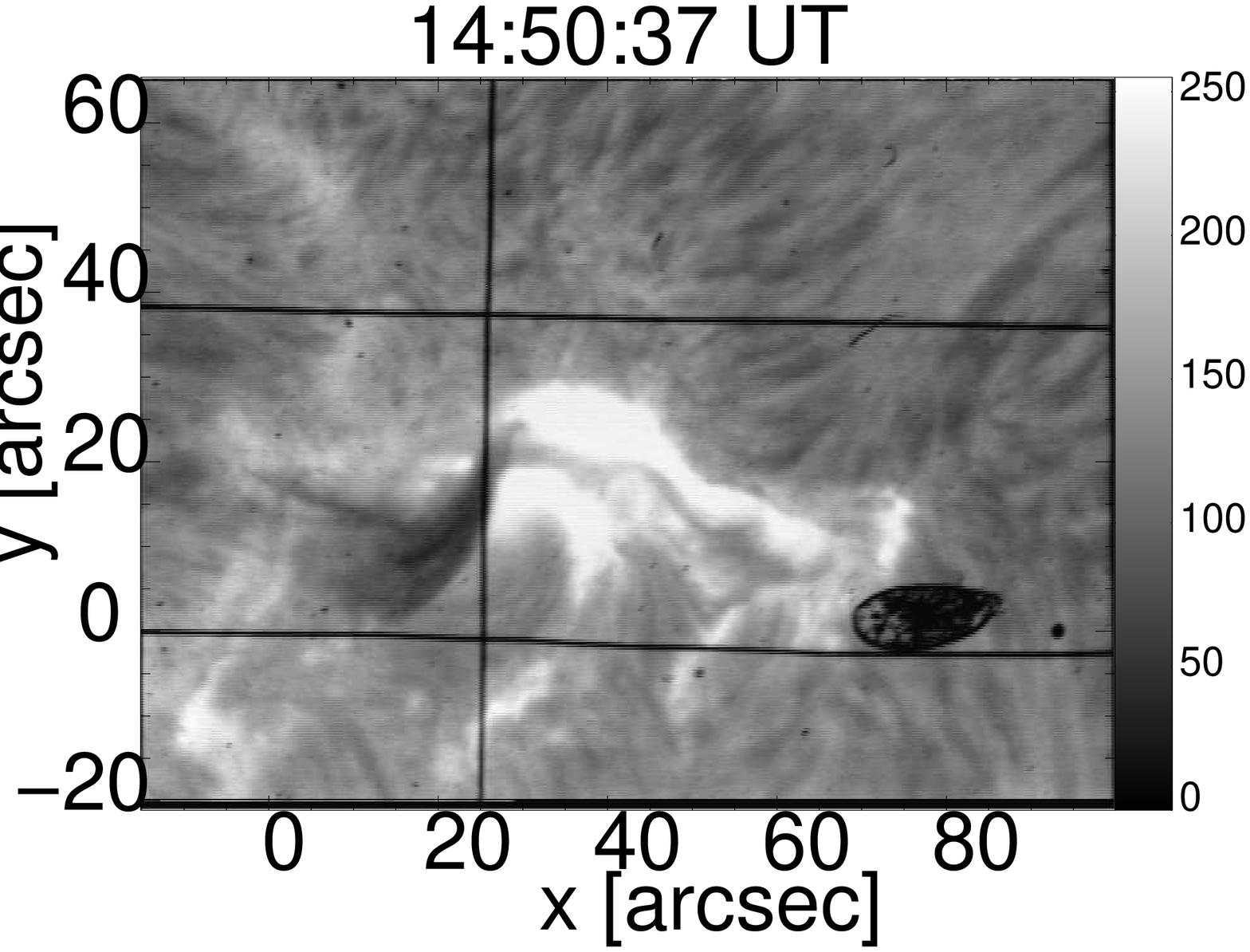}
\includegraphics[width=3.3cm,clip=true]{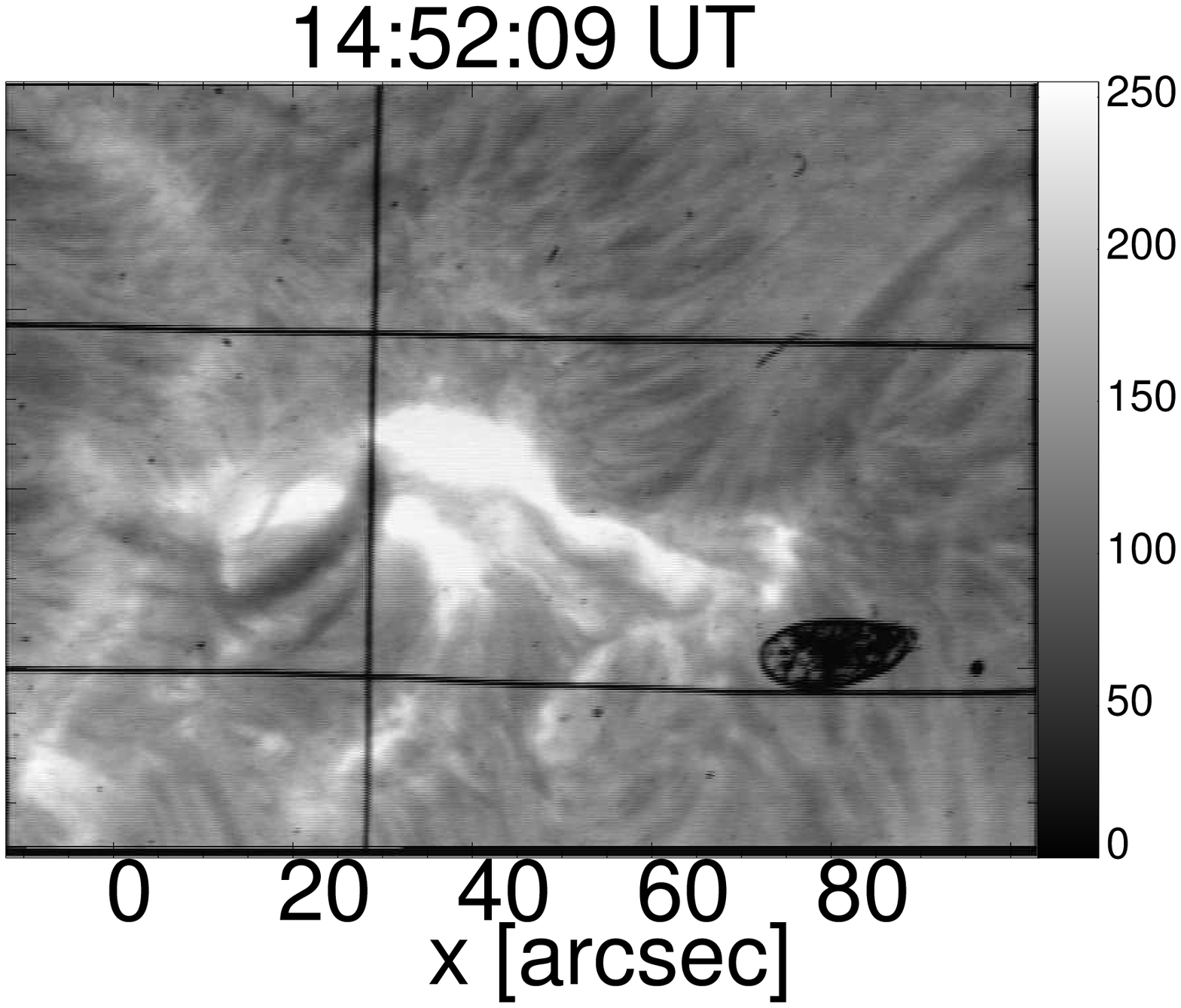}
\includegraphics[width=3.3cm,clip=true]{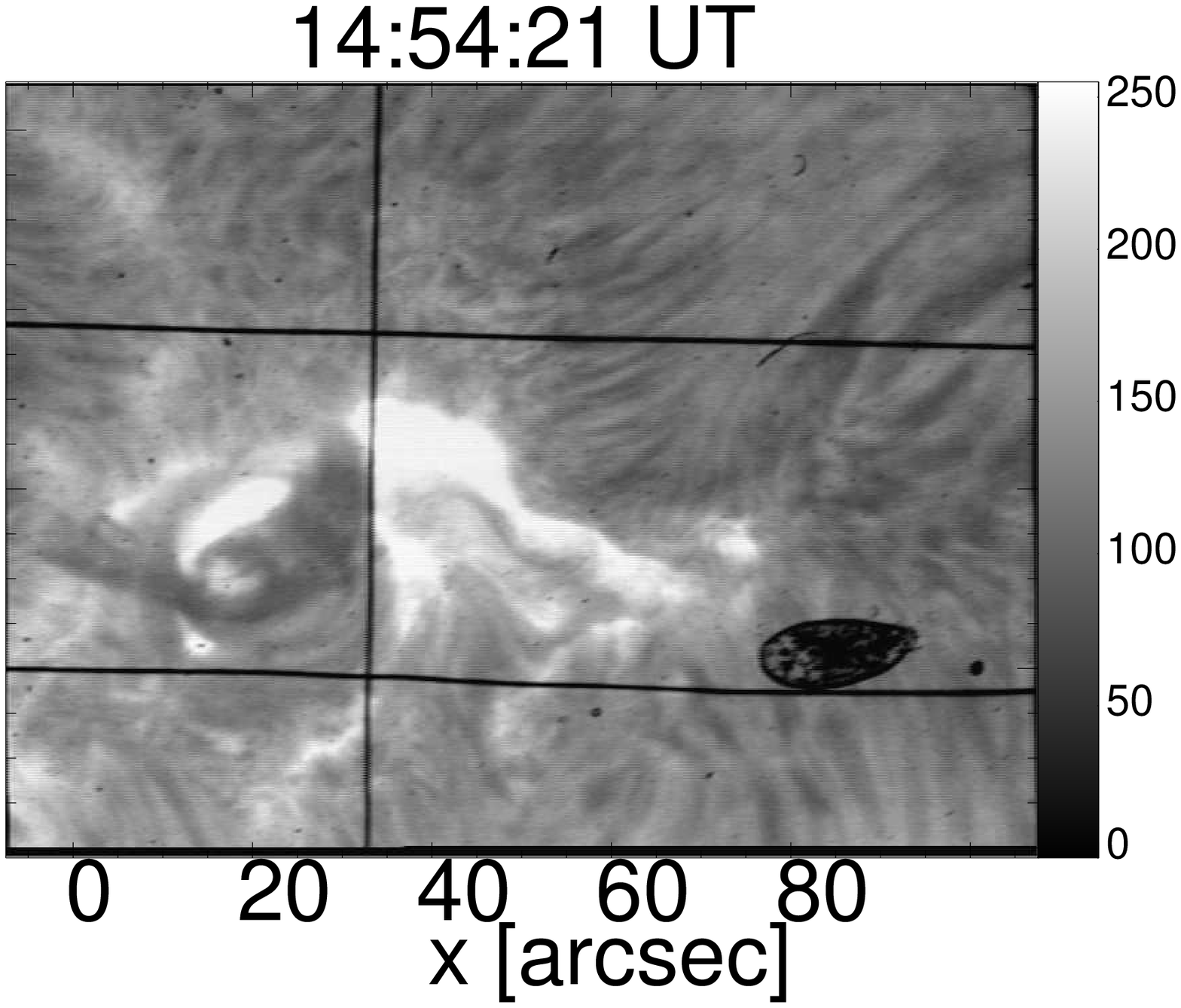}
\includegraphics[width=3.3cm,clip=true]{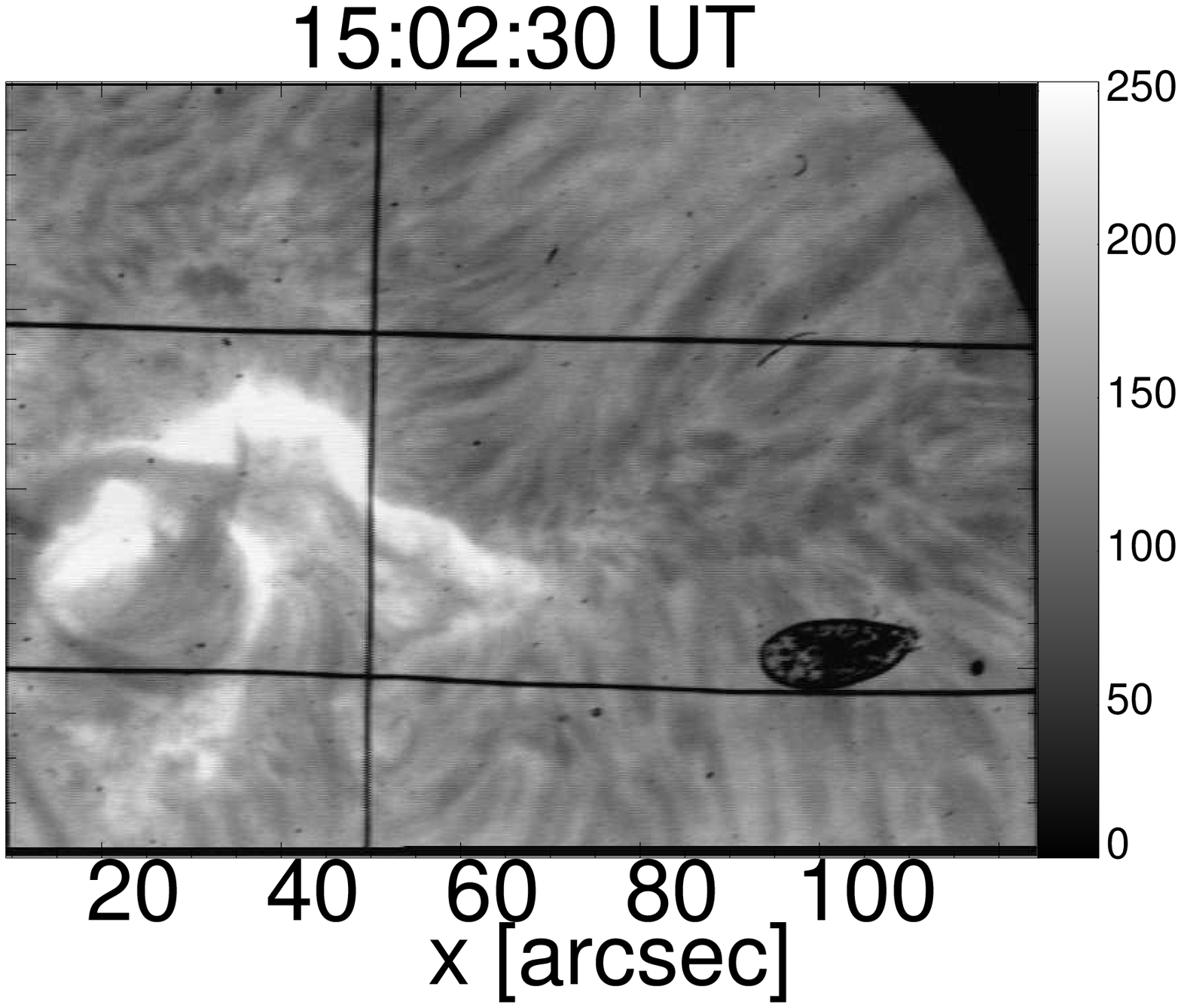}
\caption[]{\label{fig:halpha} H$\alpha$ images of the active region NOAA 10763 
obtained with the slit-jaw camera at the VTT. We can follow the evolution of 
the filament from the beginning of the observation at 14:38:33 to the end of 
the scan at 15:02:30. The drop-shaped dark feature on the lower right part of 
the image is an artefact. The temporal evolution of the filament can 
be seen in a movie attached in the on-line edition.}
\end{figure*}

To help the description of the filament's movements, we show in 
Fig.~\ref{fig:halpha} some of the simultaneous H$\alpha$ images (with the UT 
time given at the top of each) obtained with the slit-jaw camera at the Vacuum
Tower Telescope (VTT). A movie created from these and the remaining H$\alpha$
slit-jaw images is available as electronic material (halpha.wmv). The images
and the movie illustrate the very rapid evolution of the filament (darker
structure) visible at the centre of the images in between the two flare
ribbons (bright structures). The vertical line seen in the images is the slit
of the spectrometer while the two hairlines (horizontal lines) delimitate the
field of view of the spectrometer. The black spot in the lower-right corner is
an impurity on the mirror holding the spectrograph slit. It does not affect our 
spectropolarimetric observations since it is only visible in the slit-jaw 
images. When we started our scan (upper leftmost image) a part of the filament 
material had already been ejected southward after the flare began at 14:34
(according to the GOES data). The filament initially moves before starting to 
rotate, possibly after a major reconnection event. The slit moves along the 
filament during the initial rising phase, but has just reached the right edge 
of the filament and moved on to the flare ribbons at the moment that it starts 
to display a rotating motion. Looking carefully at the movie, the rotation
takes the form of a rolling motion (i.e. rotation around the axis of the flux
rope) associated with filament material loosening itself from the main body of 
the filament and moving off to the sides. The impression we get from the movie
is of an unwinding rope, of which one end has been released. Rolling motions
in filaments have earlier been noticed by, e.g., 
\citet{martin2,panasenco,suvanba}. However, as indicated earlier, the recorded
spectra are not expected to be influenced by the rotation. The filament does
not erupt completely and we could not find evidence that it was associated
with a coronal mass ejection.

The analysed data were recorded with the Tenerife Infrared Polarimeter TIP-II 
\citep{collados} mounted at the 70~cm aperture Vacuum Tower Telescope (VTT) at 
the Teide observatory in Tenerife. The filament was scanned in steps of 0.35" 
perpendicular to the slit orientation ($180.51^\circ$ with respect to the 
solar N-S direction), from 14:38:29 to 15:02:26 UT, providing a map of the 
region $36.5\times25$~Mm$^2$ in size. 

The reduction of the data has been described in Paper I. There we showed that 
the observed He Stokes profiles display a remarkably wide variety of shapes, 
and most of the profiles in the filament show very broad Stokes $I$
absorptions and complex and spatially variable Stokes $V$ signatures. The 
inversion of the profiles revealed evidence of multiple unresolved atmospheric 
blue- and redshifted components of the \ion{He}{i} lines within a single 
resolution element ($\sim 1$~arcsec), with supersonic velocities of up to 
$\sim 110$~km~s$^{-1}$. Up to five different atmospheric components were found 
in the same profile, distinguished by their Doppler shifts, which generally 
differ by more than the sound speed. Note that each component is associated
with a magnetic field. We demonstrated in Paper I that even these complex
profiles can be reliably inverted. 

The atmospheric parameters for the individual He components were retrieved by 
inverting the Stokes profiles, using the numerical code HeLIx+ 
\citep{lagg,lagg2}. The nine free parameters tuned by the code to fit the 
observed Stokes profiles were: the magnetic field strength $B$ and direction 
(inclination angle $\gamma$ and azimuthal angle $\chi$), the line-of-sight 
velocity $v_{LOS}$, the Doppler width $\Delta\lambda_D$, the damping constant 
$\Gamma$, the ratio of the line centre to the continuum opacity $\eta_0$, the 
gradient of the source function $S_1$, and the filling factor $\alpha_i$ that 
defines the contribution of a given atmospheric component $i$ to the total 
observed He profile. The sum of the filling factors of all atmospheric 
components within a resolution element is required to be unity, 
$\sum_i\alpha_i=1$. We found that the most stable fits to the observed Stokes 
profiles were obtained by coupling the magnetic field strength $B$ between the 
various Doppler-shifted He components, but leaving the two angles, $\gamma$ and 
$\chi$, of each line free. We also coupled the gradient of the source function
$S_1$ between the He atmospheric components. Coupling the field strength in
the upper chromosphere is reasonable, since, as we go higher in the solar
atmosphere, the field becomes more homogeneous in strength and more diverse in
direction \citep[e.g.][]{solanki2006}. 

As explained in Paper I, the spectral region covered by the He absorption 
signature of strongly broadened or shifted profiles contains four photospheric 
lines (\ion{Si}{i} at 10827.09~{\AA}, \ion{Ca}{i} at 10829.27 and 
10833.38~{\AA} and \ion{Na}{i} at 10834.85~{\AA}). We had to fit these lines 
as well to obtain unbiased fits of the \ion{He}{i} absorption, which often 
blends with some or all of these lines within the activated filament. Each 
photospheric spectral line was assigned its own
atmospheric component and the spectropolarimetric inversions were done by
coupling the magnetic field vector ($B$, $\gamma$ and $\chi$) and the gradient
of the source function $S_1$ between the photospheric lines, assuming that a
single set of values approximately serves all four lines. This is consistent 
with the assumption that the four photospheric lines are formed at similar
heights and can be represented by a single atmosphere. Among the photospheric
lines, the \ion{Si}{i} line is the strongest one with the best signal-to-noise
ratio and shows clear signatures in the Stokes profiles. Changes in the
atmospheric parameters will have a larger influence on the Stokes profiles of
the Si line than on the weaker photospheric lines. Therefore, the photospheric
parameters are mainly determined by the shape and strength of the Si Stokes
profiles. To obtain a good fit to the strong photospheric \ion{Si}{i} line, we
had to consider two atmospheric components: a magnetic component and a
field-free stray-light component. Only this combination could satisfactorily
reproduce both the line core and wings of the \ion{Si}{i} line. We refer to
Paper I for further details of the observations and the inversions. 

For all the inverted profile, the $v_{LOS}$ of each atmospheric component was 
well retrieved with a small error. The inclination angle $\gamma$ was
retrieved with a somewhat larger error. Nonetheless the inclinations of the
different components of the magnetic field can often be well
distinguished. The error bars on the azimuth angle, $\chi$, were instead quite
large, and it is often impossible to distinguish between the azimuths of the
different magnetic components. This is mainly because of the complexity and
the noise in the $Q$ and $U$ Stokes profiles. The information in the
observations for retrieving the azimuth is limited. We therefore refrained
from conclusions about the azimuthal direction of the magnetic field in the
filament. Because of the unreliably determined magnetic azimuth we do not
attempt to convert the magnetic vector from the observer's frame of reference
to local solar coordinates. Note that because Stokes $V$ is generally much
stronger than $Q$ and $U$, $B$cos$\gamma$ is probably the most reliably
determined quantity.

\section{Results}\label{sec:results}

\begin{figure}
\centering
\includegraphics[clip=true,width=9cm]{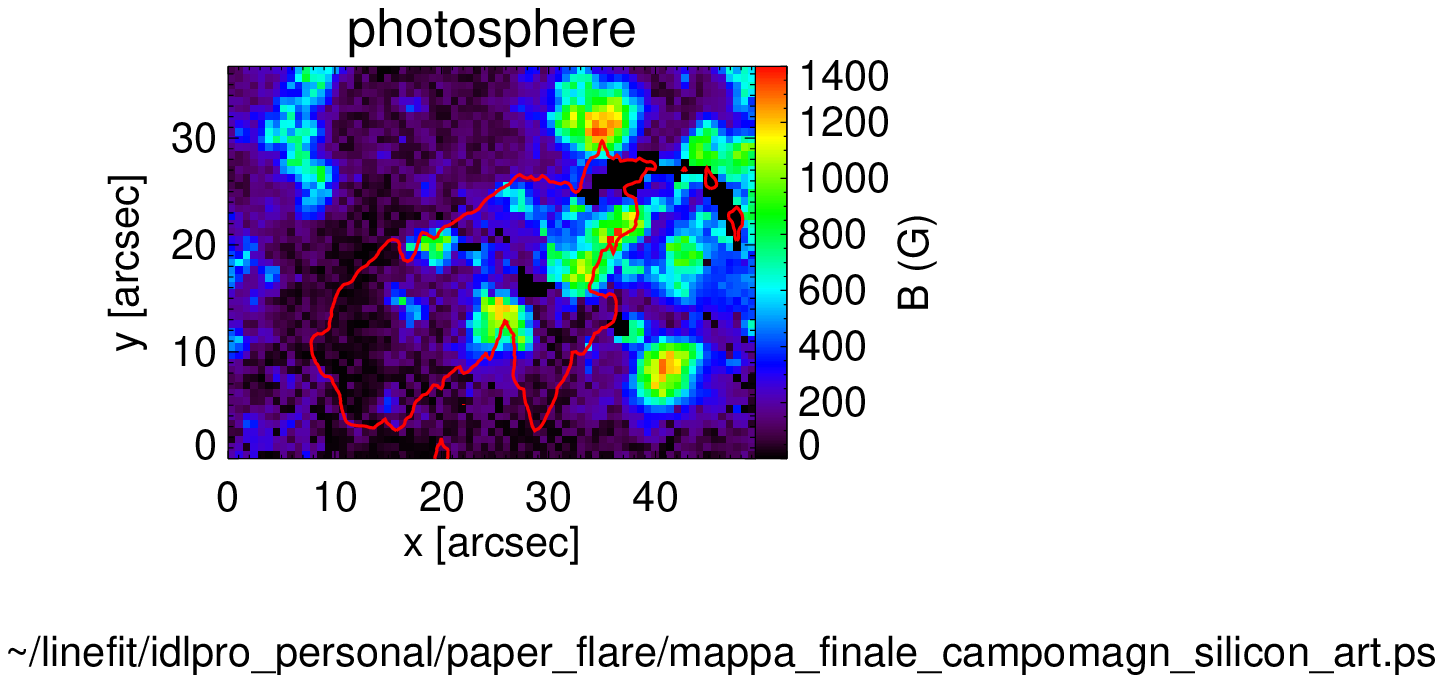}
\includegraphics[clip=true,width=9cm]{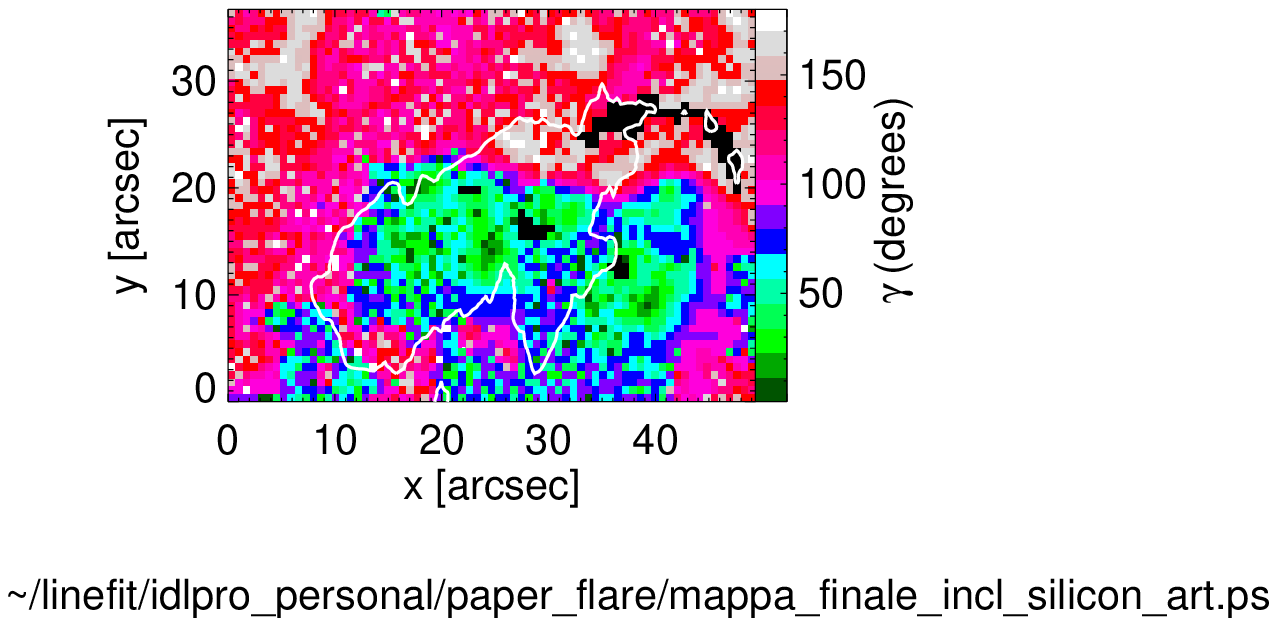}
\includegraphics[clip=true,width=9cm]{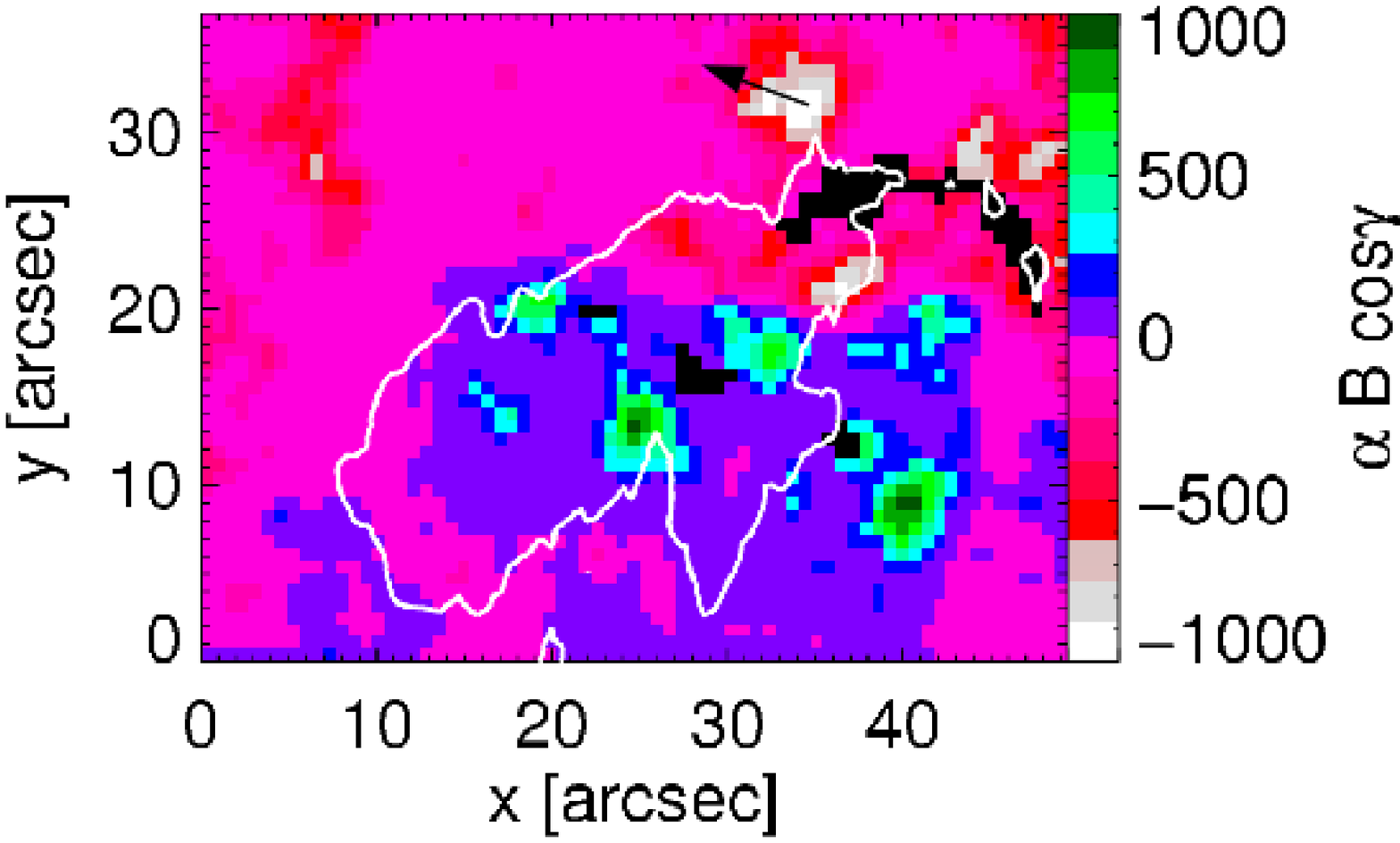}
\caption[]{\label{fig:incsilicon} Upper panel: Map of the photospheric magnetic 
field strength determined from the photospheric lines. Middle panel: Map of 
the photospheric magnetic field inclination with respect to LOS. Lower panel: 
Map of the photospheric flux density. Red and white contours 
outline the border of the filament. The black arrow in the lower panel 
indicates the direction towards solar disc centre.}
\end{figure}

Once all the profiles at each pixel position of the map are inverted, it is 
possible to create maps of the more reliably retrieved atmospheric parameters
in order to obtain a picture of the magnetic structure of the observed 
filament during its active phase. 

First, we analyse the values obtained from the photospheric lines. 
Fig.~\ref{fig:incsilicon} displays maps of the magnetic field strength (upper 
panel), the magnetic field inclination angle with respect to the LOS (middle 
panel) and the flux density ($\alpha B$cos$\gamma$, lower panel). Multiplication by the filling
factor $\alpha$, that weights the contribution of the magnetic component to
the total profile of a photospheric line (see Sect.~\ref{sec:obs}), ensures
that we have the most robustly obtained quantity. 

\begin{figure*}
\centering
\includegraphics[clip=true,width=6.8cm]{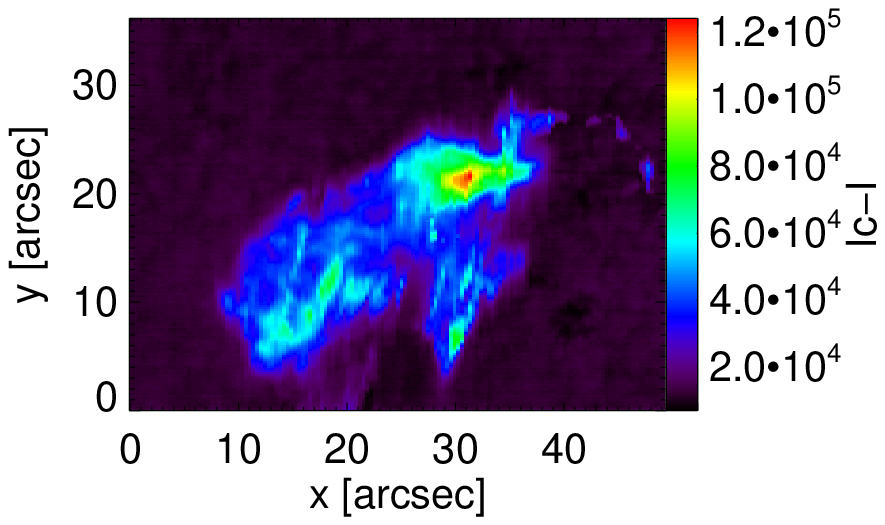}
\includegraphics[clip=true,width=6.8cm]{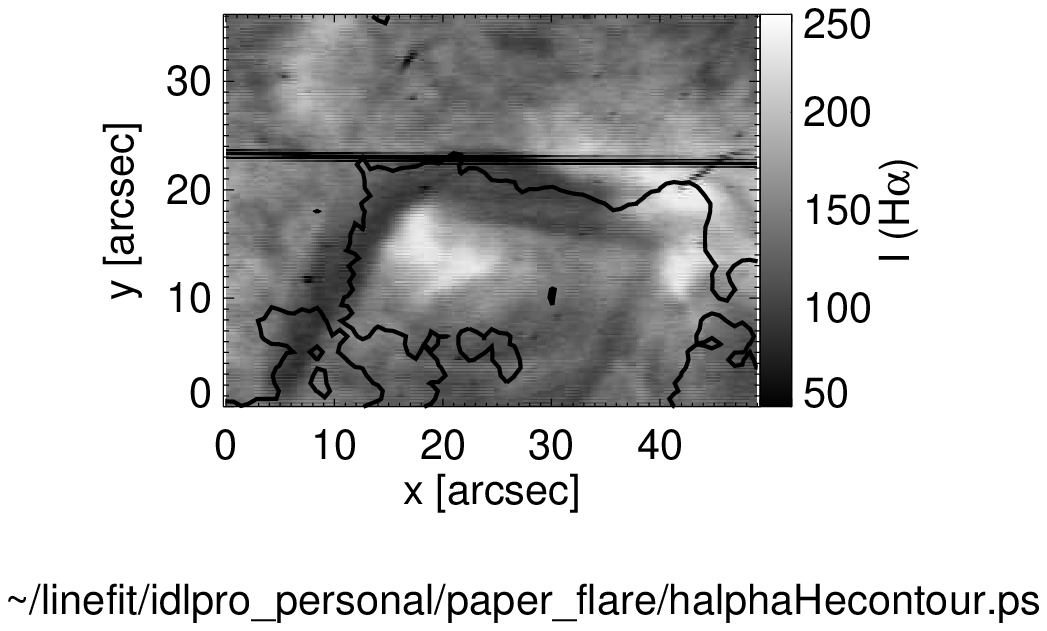}
\caption[]{\label{fig:intensitysub} Left: He intensity map obtained by 
integrating $I_\mathrm{c}-I$ over the wavelength covered by the He absorption 
(10828-10835.7 \AA). The intensity contribution of the photospheric and 
telluric spectral lines present in the chosen wavelength range has been 
subtracted. Right: H$\alpha$ image obtained with the slit-jaw camera at the 
VTT at 14:22:33 UT. The contour line indicates the position of the inversion 
line of the magnetic field in the photosphere, i.e. where $\gamma=90^\circ$ 
(see Fig.~\ref{fig:incsilicon}).}
\end{figure*}

The contour line (red or white line) in Fig.~\ref{fig:incsilicon} marks a
total line absorption of $I^*=\int(I_\mathrm{c}-I) \mathrm{d}\lambda$ equal to
$2.0 \times 10^4$~counts, where the integration runs over the wavelength range 
covered by the He absorption (10828-10835.7 \AA). The intensity contribution 
of the photospheric and telluric spectral lines present in the chosen 
wavelength range has been subtracted in order to isolate the chromospheric 
absorption. The same contour is superimposed also on many of the following 
maps. It marks reasonably well the boundary of the filament, as seen in
\ion{He}{i} 10830~{\AA}. This can be judged from the left panel of
Fig.~\ref{fig:intensitysub}, in which $I^*$ is plotted. Note that this outline
traces the evolving filament as it was sampled by the moving slit. The average
value of $I^*$ inside the filament is $\approx 3.5$ times higher than the
average value calculated in the quiet region around the filament. The maximum
value of the He absorption observed in the filament is $\approx 10$ times
bigger than the He absorption in the quiet region. 

In the middle panel of Fig.~\ref{fig:incsilicon}, we can clearly identify the 
position of the PIL where the magnetic field vector changes its polarity from 
outward (towards the observer, inclination angle close to $0^\circ$; green and 
blue colours) in the southern part of the scan to the opposite polarity 
(inclination angle close to $180^\circ$; pink and red) in the northern part. 
Figure~\ref{fig:intensitysub} (right panel) displays an H$\alpha$ image 
obtained with the slit-jaw camera at the VTT at 14:22:33 UT, prior to our scan 
and prior to the flare eruption. The filament mainly overlies the photospheric 
polarity inversion line (contour line drawn at $\gamma=90^\circ$, determined
from the middle panel of Fig.~\ref{fig:incsilicon}). The horizontal black line 
is one of the hairlines marking the field-of-view of the TIP-II
instrument. The filament contour in \ion{He}{i} (obtained from the scan 
lasting from 14:42:16 to 14:56:54) only partly straddles the inversion line
(see Fig.~\ref{fig:incsilicon}), unlike the filament seen in H$\alpha$ at
14:22:33 (Fig.~\ref{fig:intensitysub}, right panel). The strong He absorption
at least in the first part of the scan is clearly situated only above one
polarity of the photospheric field, suggesting that the filament (or the flux
rope structure) was evolving rapidly after activation by the flare. The movement
of the material away from the photospheric inversion line is also deduced from
the simultaneous H$\alpha$ images obtained with the slit-jaw camera at the VTT
(see Sect.~\ref{sec:obs}).

\begin{figure*}
\centering
\includegraphics[clip=true,width=6.8cm]{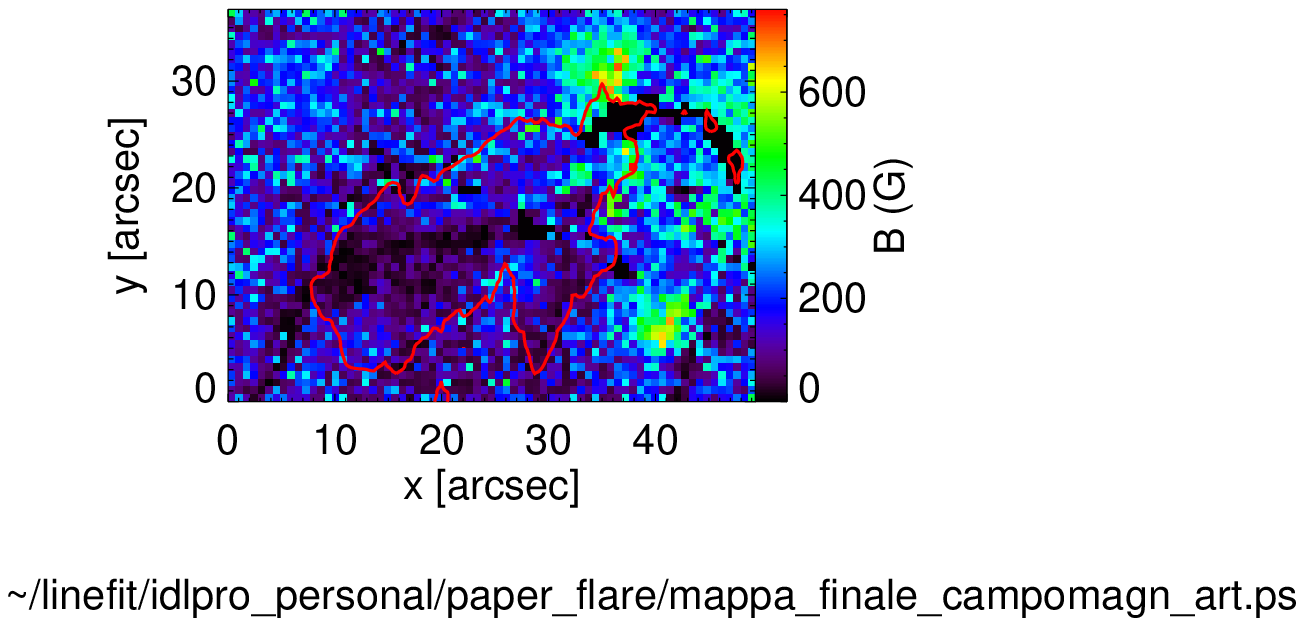}
\includegraphics[clip=true,width=6.8cm]{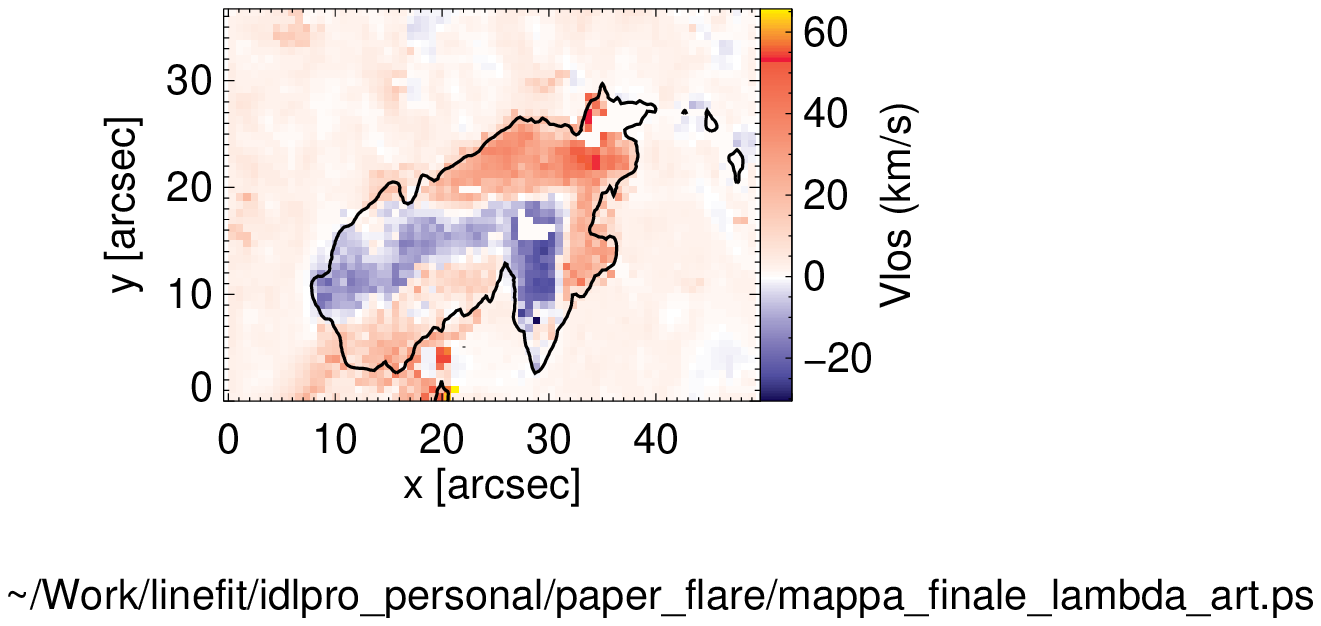}
\caption[] {\label{fig:campomagnchromo} Left: Map of the chromospheric magnetic 
field strength. Right: Map of the mean value of the retrieved $v_{LOS}$ for 
the He components weighted with the respective filling factor $\alpha$. Red 
and black contours outline the position of the filament as sampled by the
spectropolarimeter's scan. They are defined as 
$I^*=\int(I_\mathrm{c}-I) \mathrm{d}\lambda=2\times10^4$ counts (see left 
panel of Fig.~\ref{fig:intensitysub}).}
\end{figure*}

Figure~\ref{fig:campomagnchromo} (left panel) displays the map of the retrieved 
chromospheric magnetic field strength. As explained in Sec.~\ref{sec:obs}, we 
obtained the best fits to the observed Stokes profiles by coupling the 
magnetic field strength $B$ between the He components. The field strength 
values are generally lower than 400~G inside the region of strong He 
absorption, which is indicated by the red contour line. Stronger field regions 
correspond to the positions of small pores visible in the continuum intensity 
map (Fig.~1 in Paper I), and other prominent locations of strong field in the 
photosphere (Fig.~\ref{fig:incsilicon}). The average field strength in the 
filament $\sim 119$~G is lower than its average value in the region 
around it $\sim 190$~G. This difference probably reflects the fact that the
\ion{He}{i} line becomes optically thick inside the
filament and is formed considerably higher than the surrounding atmosphere.

As a quick look, helpful for the He data interpretation and the discussion of
the results that will be done in the next section, we show the overall
chromospheric velocity pattern (Fig.~\ref{fig:campomagnchromo}, right panel). 
We calculated the weighted mean value of $v_{LOS}$ for each spatial pixel. The 
averaging was done over all atmospheric components deduced from the He-triplet 
inversions, weighted with their respective filling factors $\alpha$. The
downflows are mainly located at the edges of the filament (possibly
corresponding to the footpoints), with the fastest ones at the right 
edge, while the upflows are mainly located in the body of the filament.  
As we already noted in Sect.~\ref{sec:obs}, the rolling of the filament happens
after the slit has left the filament. This is consistent with the absence of 
any preferred velocity pattern indicative of a rolling filament in our 
resulting velocity maps.

\begin{figure*}
\centering
\includegraphics[clip=true,width=2.9cm]{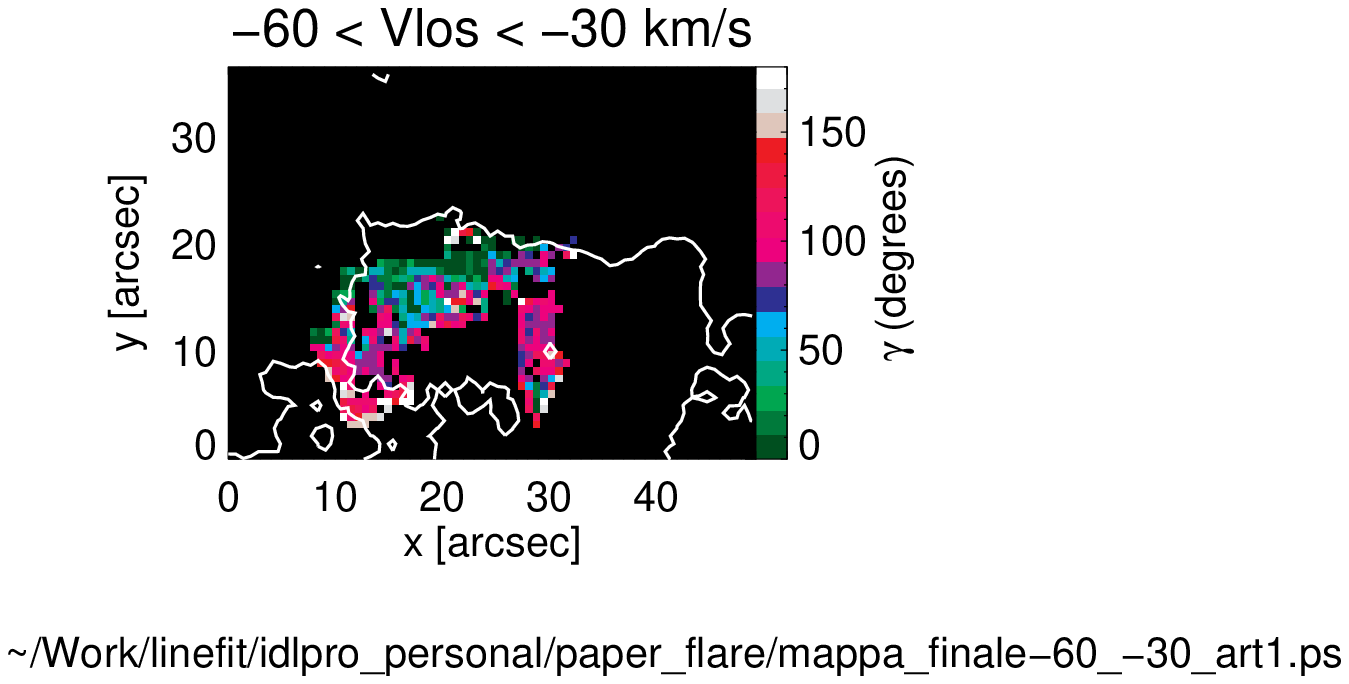}
\includegraphics[clip=true,width=2.35cm]{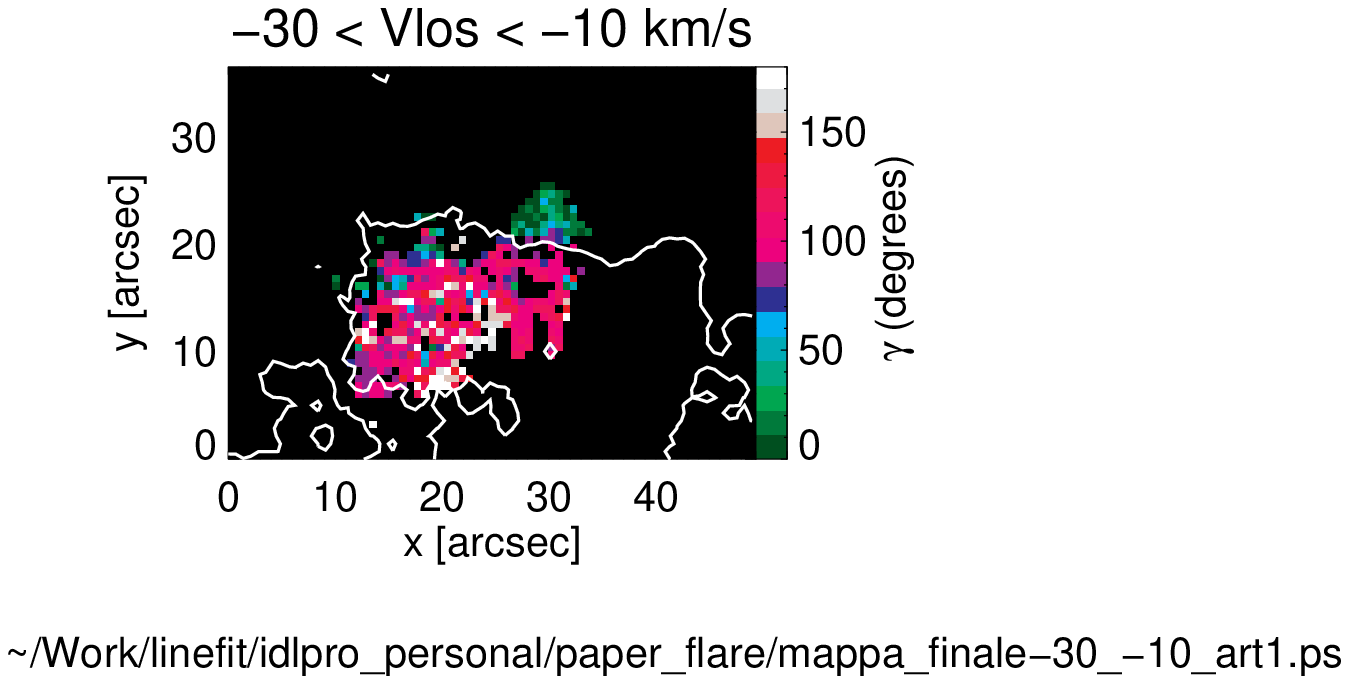}
\includegraphics[clip=true,width=2.35cm]{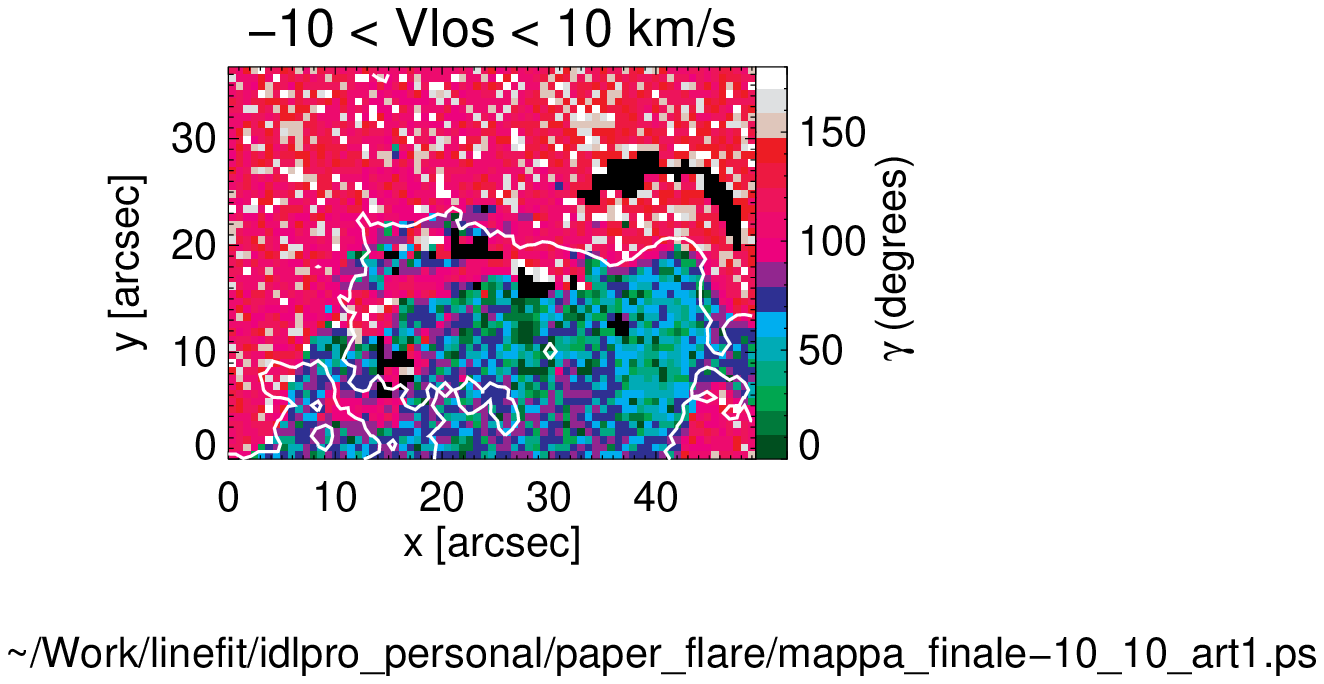}
\includegraphics[clip=true,width=2.35cm]{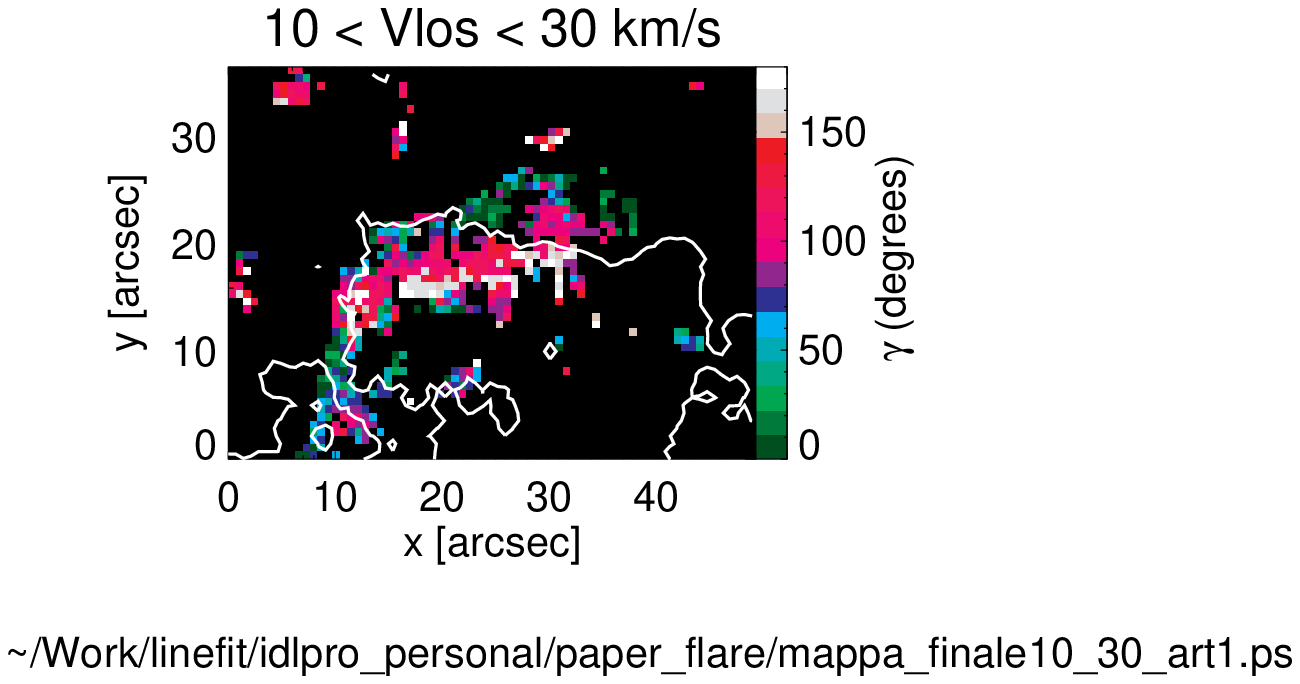}
\includegraphics[clip=true,width=3.1cm]{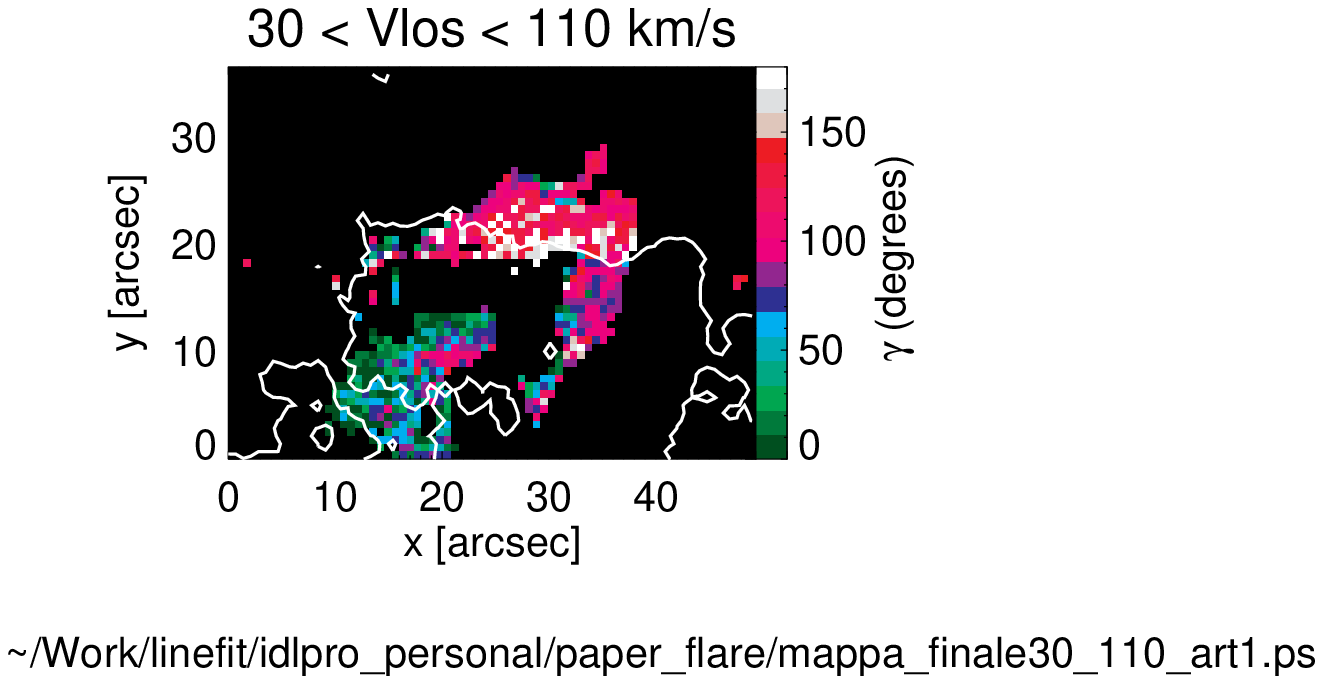}
\hspace{2.0cm}
\includegraphics[clip=true,width=2.9cm]{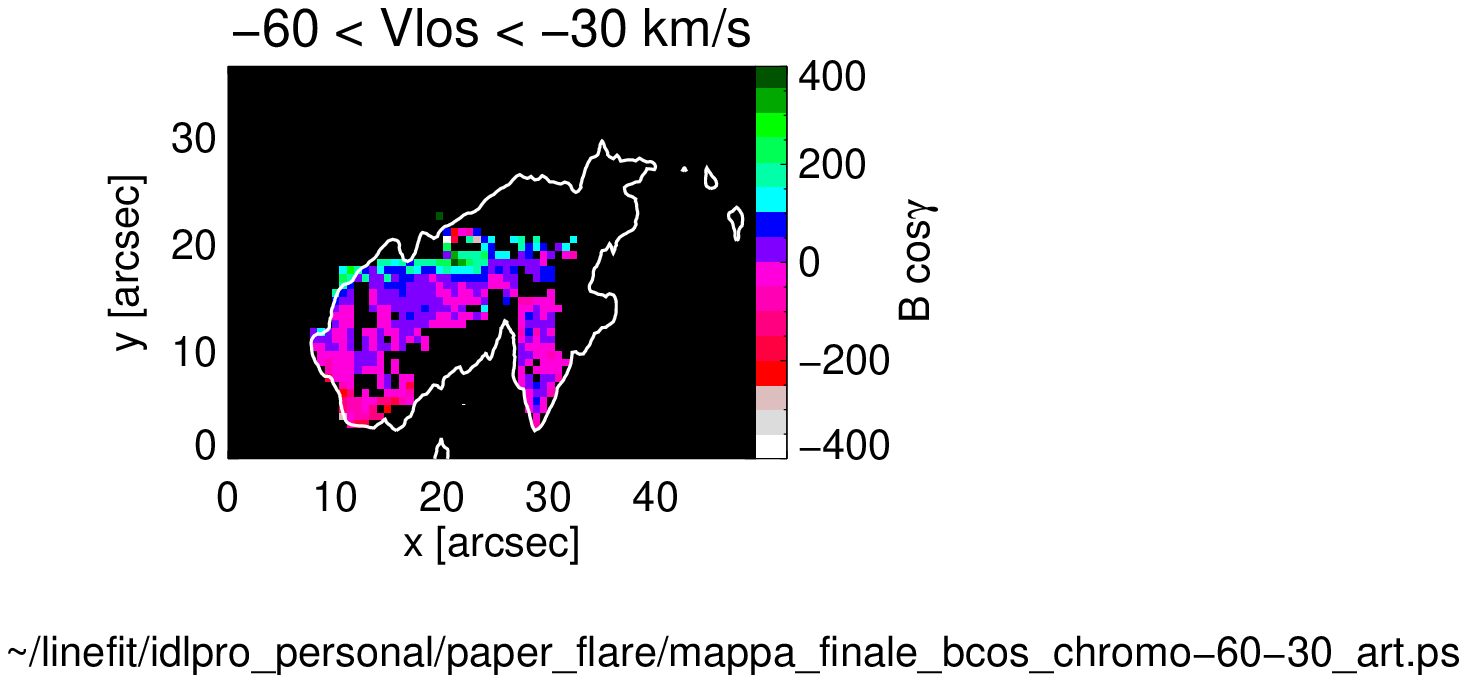}
%\hspace{0.1mm}
\includegraphics[clip=true,width=2.4cm]{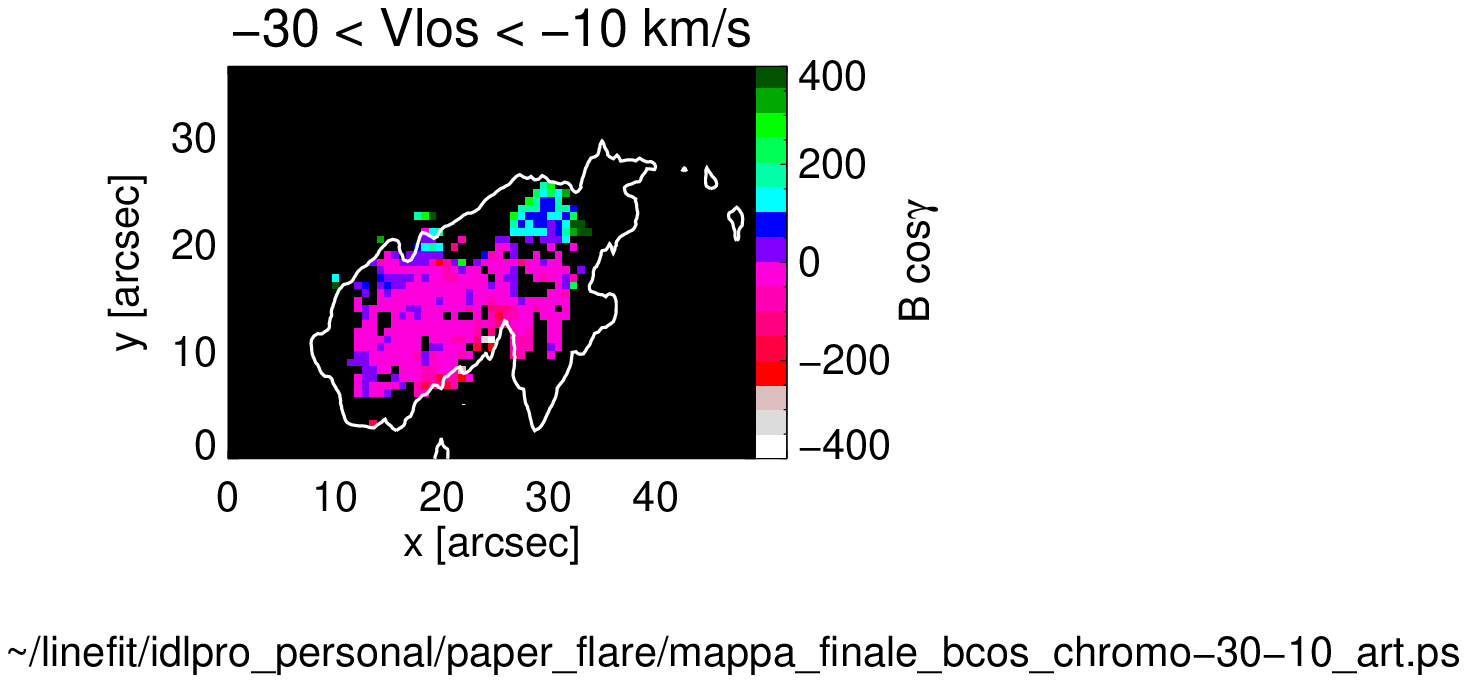}
\includegraphics[clip=true,width=2.4cm]{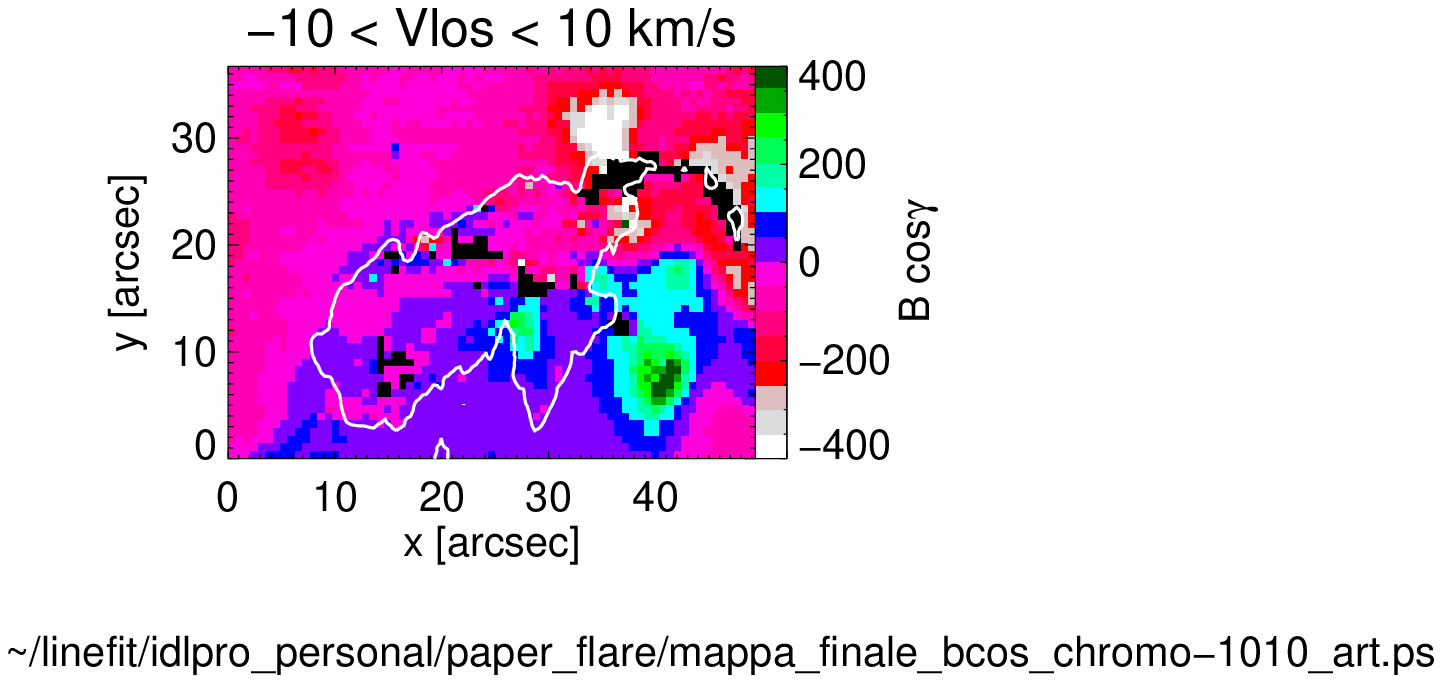}
%\hspace{0.1mm}
\includegraphics[clip=true,width=2.4cm]{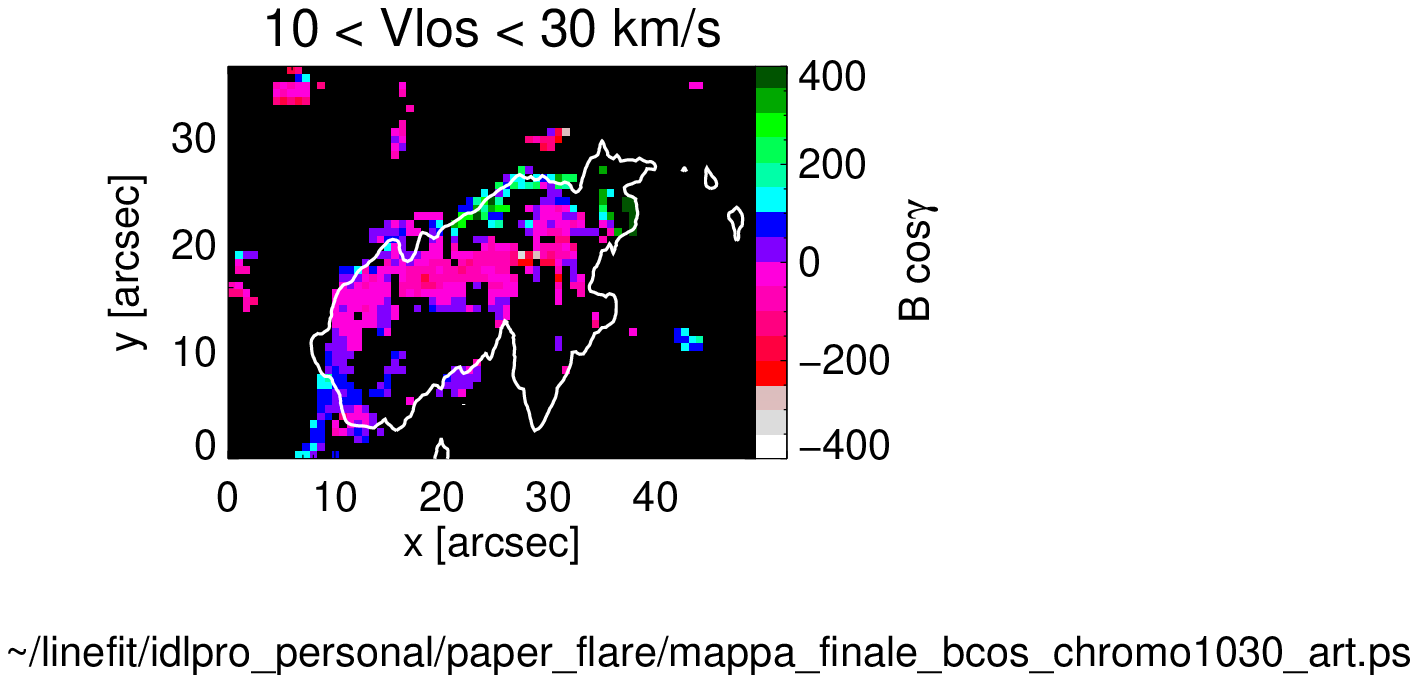}
\includegraphics[clip=true,width=3.1cm]{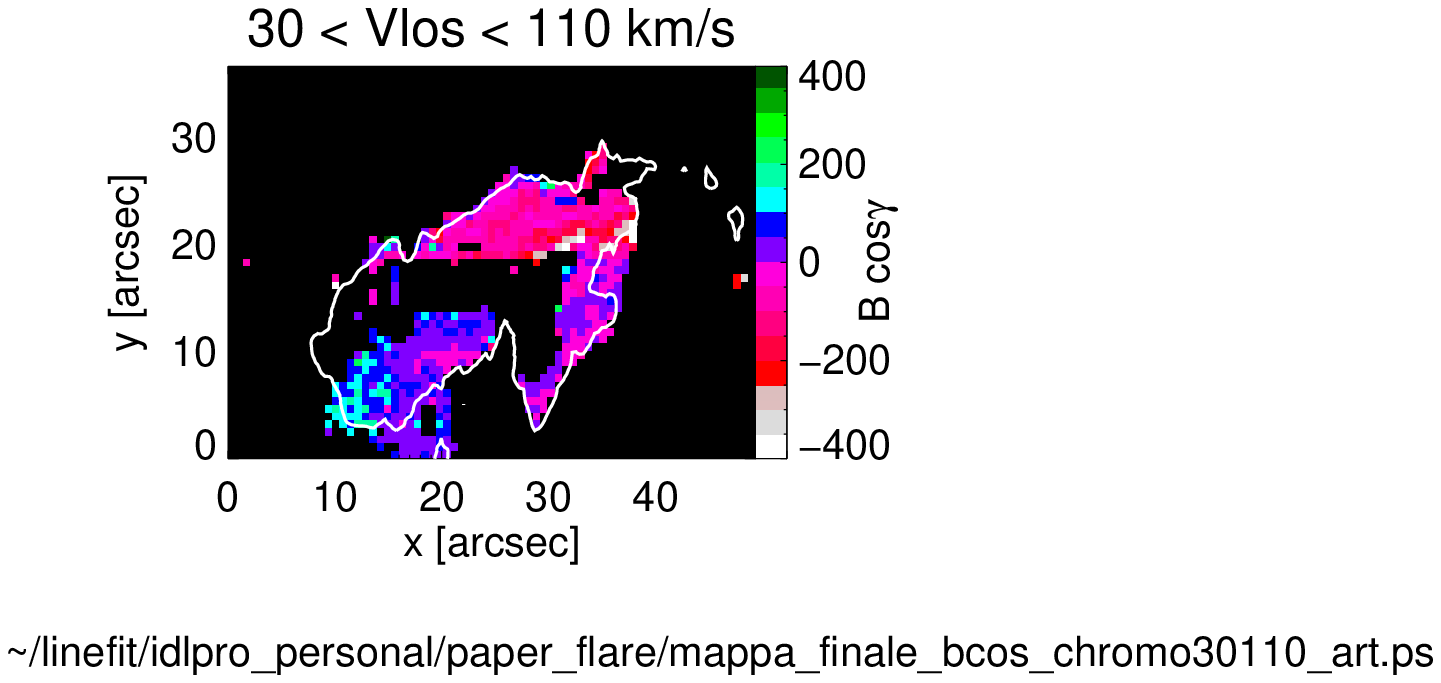}
\caption[]{\label{fig:inc6030} Maps of the retrieved magnetic field 
inclination ($\gamma$, upper panels) and of the LOS magnetic field component 
($B$cos$\gamma$, lower panels) for the He components with retrieved LOS 
velocity in the range $-60<v_{LOS}<-30$~km~s$^{-1}$, 
$-30<v_{LOS}<-10$~km~s$^{-1}$, $-10<v_{LOS}<10$~km~s$^{-1}$, 
$10<v_{LOS}<30$~km~s$^{-1}$, and $30<v_{LOS}<110$~km~s$^{-1}$. The contour 
line in the upper panels indicates the position of the PIL in the
photosphere. The contour line in the lower panels outlines the region of strong 
He absorption.}
\end{figure*}

Maps of the retrieved chromospheric magnetic field inclination with respect to 
the LOS are shown in the upper panels of
Fig.~\ref{fig:inc6030}. Inclinations for five ranges of LOS velocity are
separated to distinguish between the inclinations retrieved from the different
He components. The chosen $v_{LOS}$ ranges are: $-60<v_{LOS}<-30$,
$-30<v_{LOS}<-10$ (blueshifted components), $-10<v_{LOS}<10$ (components
nearly at rest), $10<v_{LOS}<30$, and $30<v_{LOS}<110$~km~s$^{-1}$ (redshifted
components). Magnetic inclinations associated with these $v_{LOS}$ ranges are
plotted from left to right in the upper panels of Fig.~\ref{fig:inc6030}. If
in one pixel multiple He components with $v_{LOS}$ in a given range coexist,
then we plot the inclination of the component with the highest filling
factor. The contour line indicates the position of the apparent PIL in the
photosphere, i.e. where $\gamma=90^\circ$ (see Fig.~\ref{fig:incsilicon}).
The lower panels display the values of the LOS magnetic field in the
atmospheric components within the different velocity bands, while the
contour line outlines the region of strong He absorption.

We first point to the general distribution of the signal in the different
frames. This shows the locations of the regions with gas within the velocity
ranges covered by the individual panels. Clearly, outside the filament only
gas nearly at rest is present, with the exception of a few isolated locations
of downflows less than 30~km~s$^{-1}$. There is also a difference in the
distribution of the most strongly upflowing and the most strongly downflowing
gas (compare the leftmost with the rightmost panels). Whereas the upflowing
gas tends to be located in the body of the filament, the very rapidly
downflowing gas is concentrated at its periphery. This confirms the general
conclusion drawn from the right panel of Fig.~\ref{fig:campomagnchromo}. There
is little spatial overlap between the fastest downflows and upflows.

\begin{figure*}
\centering
\includegraphics[clip=true,width=6.5cm]{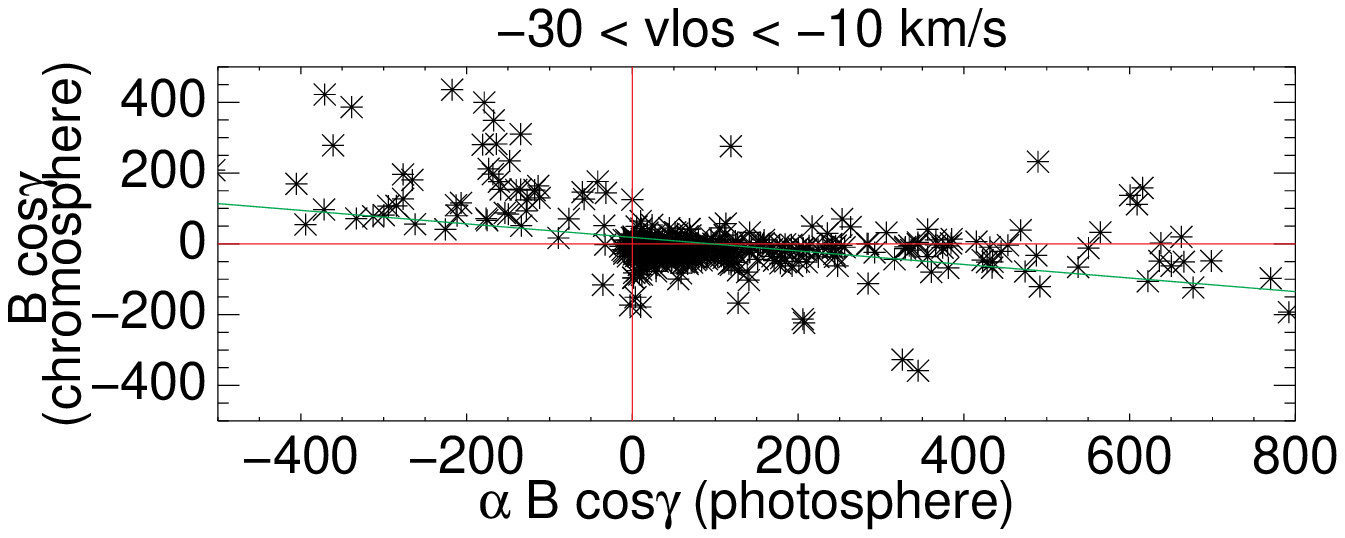}
\includegraphics[clip=true,width=6.5cm]{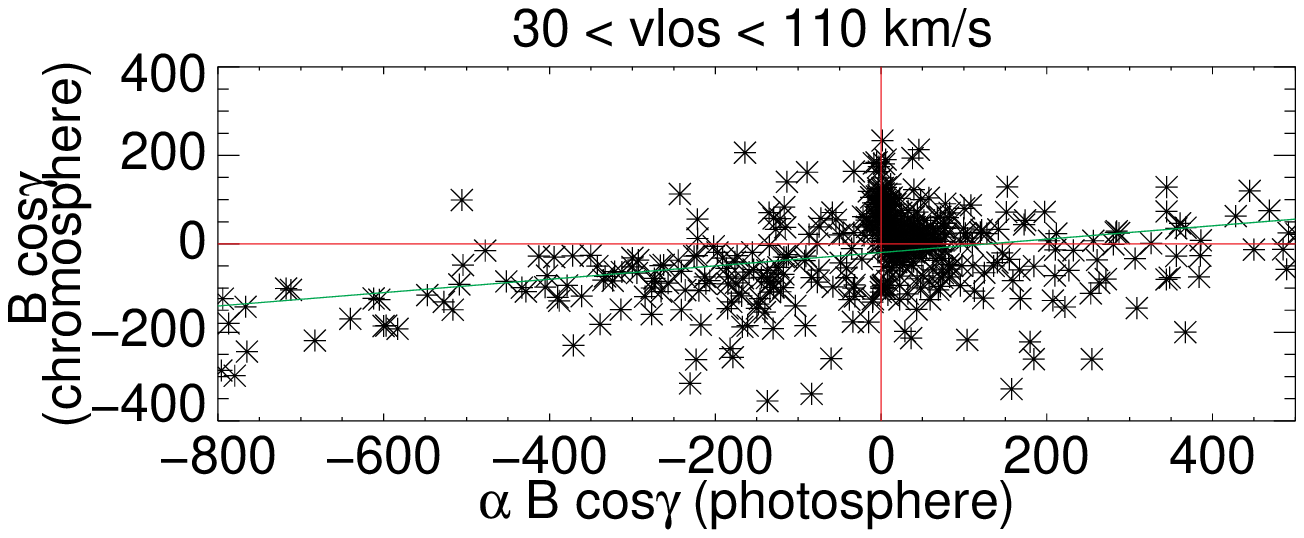}
\includegraphics[clip=true,width=6.5cm]{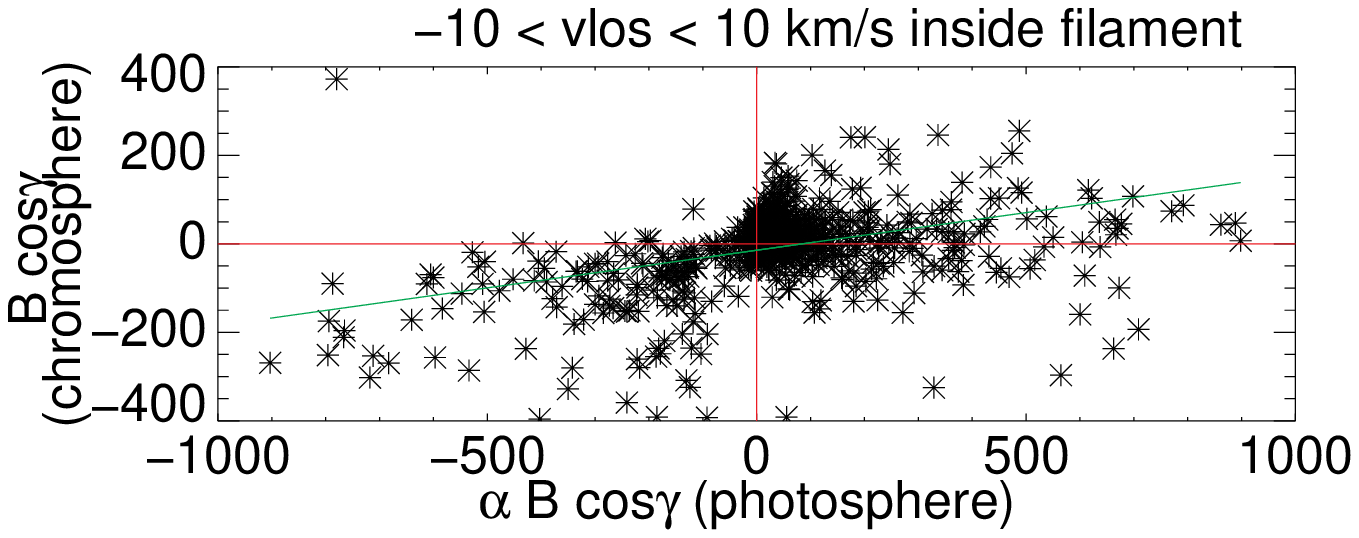}
\includegraphics[clip=true,width=6.5cm]{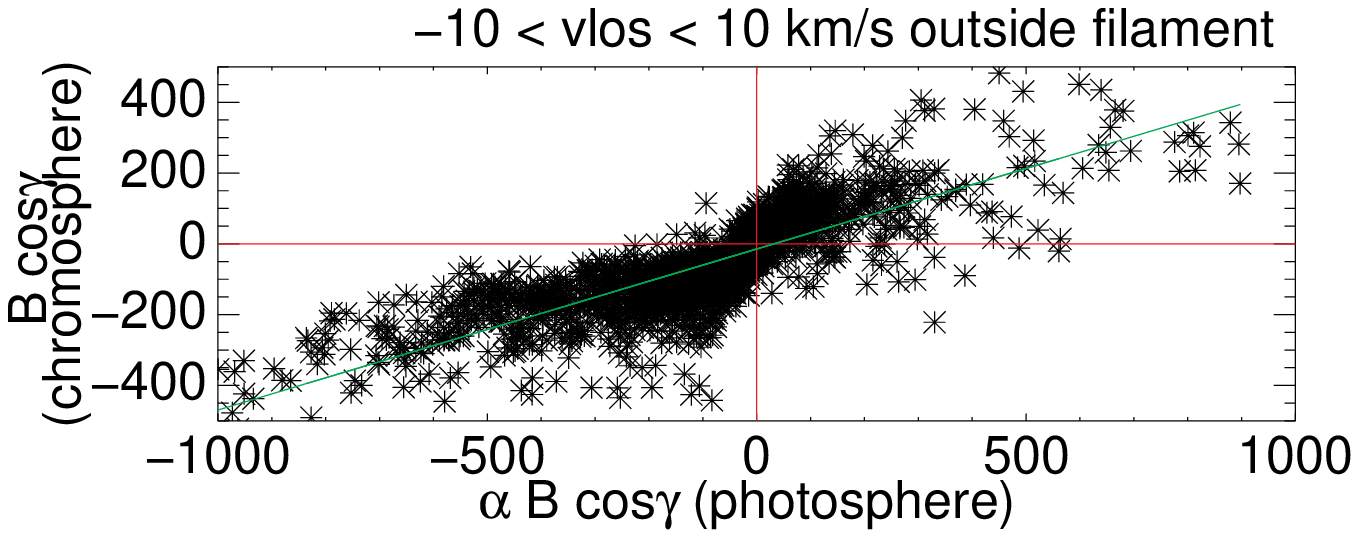}
\caption[]{\label{fig:scatter60110} LOS magnetic field component in the 
chromosphere versus the LOS magnetic field component in the photosphere for 
the He components with retrieved LOS velocity in the range
$-30<v_{LOS}<-10$~km~s$^{-1}$ (top left), $30<v_{LOS}<110$~km~s$^{-1}$ 
(top right), and $-10<v_{LOS}<10$~km~s$^{-1}$ (bottom panels). The green lines 
represent linear regressions.}
\end{figure*}

Let us next consider the maps in the first and second columns of 
Fig.~\ref{fig:inc6030}, i.e. those derived from the blueshifted He
components. These maps show a concentration of almost transversal field 
($\gamma\approx90^\circ$) between two regions of more longitudinal field lines 
with opposite polarities. The magnetic field vector 
points outward (towards the observer) in the northern part of the scan and 
inward in the southern part. The positions of the two polarities are opposite 
to the ones we retrieved in the photosphere, where the magnetic vector was
instead pointing outward in the southern part of the scan and inward in the
northern part. This behaviour is supported for the He components with 
$-30<v_{LOS}<-10$~km~s$^{-1}$ by Fig.~\ref{fig:scatter60110}, where we 
plot the LOS magnetic field component in the chromosphere versus the LOS 
magnetic field component in the photosphere for different He components 
(within specific velocity ranges). For the slower blueshifted He components
($-30<v_{LOS}<-10$~km~s$^{-1}$, top left panel), when the LOS magnetic field
component in the photosphere is positive (pointing towards the observer), it
appears to be negative in the chromosphere and vice versa (although the
relationship is less clear-cut in the other polarity, possibly due to
line-of-sight effects). The scatter plot for the range
$-60<v_{LOS}<-30$~km~s$^{-1}$ does not show a clear-cut relationship because
of the displacement of the neutral line in the chromosphere with respect to
the photosphere, which can be deduced from the top left panel of
Fig.~\ref{fig:inc6030}. Hence, the same pixel in the photospheric and
chromospheric map cannot be used to check the correspondence in
polarity. However, it is clear from the comparison of
Fig.~\ref{fig:incsilicon} with the leftmost panels of Fig.~\ref{fig:inc6030}
that whereas the positive polarity in the photosphere is mainly concentrated
on the south side of the filament, in the chromosphere the positive polarity
is concentrated mainly to the north of the negative polarity.
  
From the maps in Figs.~\ref{fig:incsilicon} and \ref{fig:inc6030}, we also 
notice that the photospheric field is more longitudinal than the chromospheric 
one for which $\gamma$ is more concentrated around the value of $90^\circ$. 
This result could reflect the situation on the Sun, but can also be an effect 
of the noise, since the signals in He are weaker than in Si. The resulting 
lower signal-to-noise ratios lead to the inversion code returning a less
vertical field from the He lines than may actually be present in the upper 
chromosphere. 

The panels of the third column of Fig.~\ref{fig:inc6030} depict the maps 
related to chromospheric gas almost at rest ($-10<v_{LOS}<10$~km~s$^{-1}$). 
Outside the region covered by the filament but also for the majority of the
profiles observed in the filament, the magnetic field maintains the same
polarities in both the photosphere and the chromosphere (see also the bottom
panels of  Fig.~\ref{fig:scatter60110}, left for the points inside the 
filament and right for the points outside). In some regions in the filament, 
however, in particular for $10"<x<25"$, $3"<y<18"$, the chromospheric magnetic 
field changes its polarity multiple times across the filament, while the 
photospheric field does not. Hence in this region the chromospheric field 
points in the opposite direction as compared to the photospheric magnetic 
field. This is not well visible in Fig.~\ref{fig:scatter60110} because
$B$cos$\gamma$ is small at these locations. 

The redshifted He components displayed in the fourth and fifth columns of
Fig.~\ref{fig:inc6030} exhibit distributions of polarities of the magnetic 
field that resemble the photospheric pattern. The field lines go from outward 
(i.e. directed towards the observer) in the southern part of the scan to the 
opposite polarity in the northern part. This behaviour is confirmed by the 
scatter plot in the top right panel of Fig.~\ref{fig:scatter60110} for the 
fastest downflows: the LOS magnetic field components in the photosphere and in 
the chromosphere show a positive correlation, although with considerable
scatter due to the mismatch of the PILs in the photosphere and
chromosphere. In the range $10<v_{LOS}<30$~km~s$^{-1}$ the qualitative
behaviour is the same as for the stronger redshifts, but (as in case of the
profiles with $-60<v_{LOS}<-30$~km~s$^{-1}$) the mismatch between the PILs in
the photosphere and the chromosphere produces a lot of scatter in a
quantitative pixel-by-pixel comparison.

\section{Discussion and Conclusions}\label{sec:conlusions}

\begin{figure*}
\centering
\includegraphics[clip=true,width=6cm]{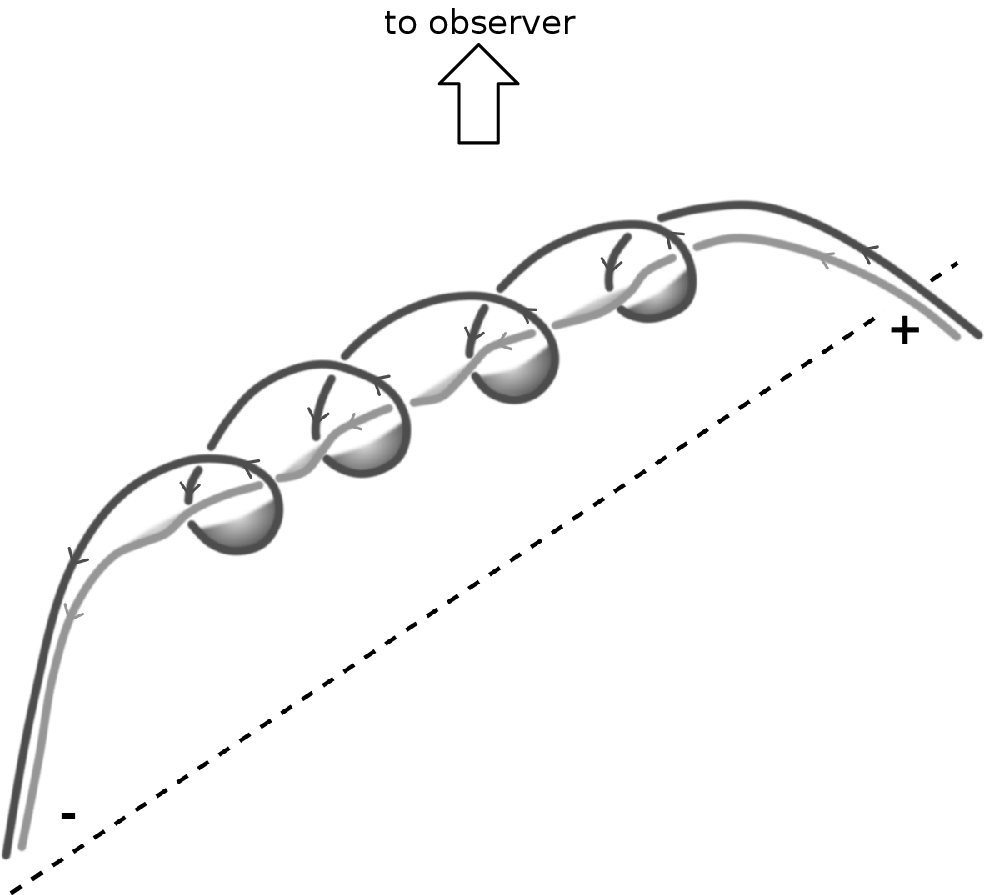}
\includegraphics[clip=true,width=6cm]{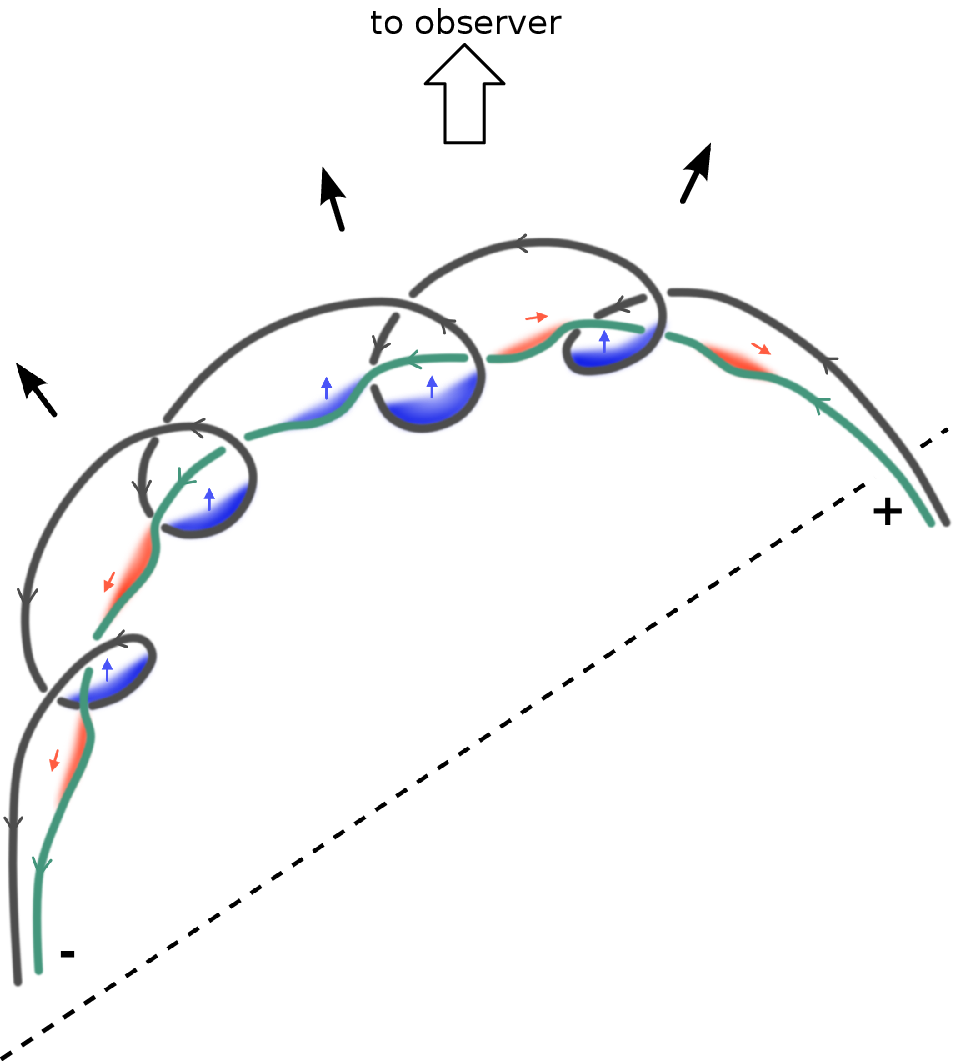}
\caption[]{\label{fig:model2} First proposed scenario. Plotted are 
two of the magnetic field lines with different degrees of twist belonging to
the flux rope underlying the filament. In the left panel, the flux rope is in 
equilibrium, while in the right panel it is rising up. The filament material is 
represented by the shadowed regions, where the colours indicate if the 
material is blue- or redshifted relative to the observer (located above the
top of the image). The dashed line is the photospheric polarity inversion line.}
\end{figure*}

We have presented the, to our knowledge, first maps of velocity and magnetic
field of an activated filament in a flaring active region. Such maps are
created in both the photosphere and the chromosphere. Up to 5 different
magnetic components are found in the chromospheric layers of the filament,
while outside the filament a single component is sufficient to reproduce the
observations. These components differ from each other in the gas flows they
harbour and in the magnetic field inclination, with different components
deduced from the same set of Stokes parameters sometimes having opposite
magnetic polarities. 

The reliability of the deduced properties of these components was carefully
discussed in Paper I. In addition, it is also suggested, at least 
qualitatively, by the fact that the resulting maps of velocity and magnetic
inclination are reasonably smooth and consistent, although individual profiles
were inverted without using the values obtained from neighbouring profiles to
guide the solution in any way (recall that the HeLIx+ code makes use of a 
genetic algorithm, which searches for the global minimum in the $\chi^2$ 
surface for each and every inverted profile). 

These results suggest that the activated filament has a rather complex
structure. However, as we shall show below, this structure is compatible
with a simple flux rope, although with a distorted one. In the following, we
combine the information obtained from the inversions of the
spectropolarimetric data to compare it with the different models of filament
support. We find best qualitative agreement of our observations with the flux
rope model. In particular, this model explains well the presence of upflows
and downflows, showing opposite magnetic polarities within one and the same
resolution element.  

Let us summarise the main results obtained from the inversions. We measure 
downflow velocities of up to 110~km~s$^{-1}$ and upflows of up to 
60~km~s$^{-1}$ along the LOS. These supersonic flows always coexist within the
same resolution element with a component of the upper chromosphere almost at
rest, or at least displaying a subsonic flow ($-10<v_{LOS}<+10$~km~s$^{-1}$). 
Upflows and downflows are both found throughout nearly the entire filament, 
with the following exceptions: At the end points of the filament (i.e. in its 
lower left and upper right corners) we see only fast downflows. They are 
associated with a relatively longitudinal field having the same polarity as
the underlying photospheric field. The central part of the filament harbours
mainly upflows (although weaker downflows are present there too). The 
blueshifted components of the \ion{He}{i} triplet are in general associated 
with more transverse field lines, although closer to the edges of the filament
the blueshifted gas is co-located with more longitudinal field showing
opposite polarities to the downflowing gas in the same pixels. I.e. the
upflowing gas is associated with opposite polarity fields compared with the 
photosphere. We interpret this magnetic field with polarities opposite 
to the photospheric, associated with upflows as signatures of the dips
harbouring stable plasma in the rising flux rope.

The downflowing gas is hence associated with more $\Omega$-shaped field lines,
while the upflowing gas appears to be carried by U-shaped field lines. How can
this complex picture be reconciled with a model of a filament? Taking the sum 
of these observations we propose the following scenario: 

Prior to its activation the filament is described by a relatively horizontal
flux rope whose photospheric foot points are located at its lower left and
upper right corners. Such low-lying filaments are typical for active regions 
\citep{zirin,martin,thb}. Once this filament gets activated, probably by
the nearby flare, it starts evolving rapidly, moving horizontally (see
H$\alpha$ images in Fig.~\ref{fig:halpha} and the movie) and in particular
rising, specially in the middle. This distorts the field. The axis of the flux
rope, which originally was relatively horizontal for most of the length of the 
filament (see the rough sketch of the filament before it was activated in the 
left frame of Fig.~\ref{fig:model2}) becomes more curved, reaching higher in 
the centre, as indicated in the right image of Fig.~\ref{fig:model2}. 
Later, possibly after a major reconnection event, the movie shows that the
filament starts to roll (maybe unwinding). Unfortunately, we do not have 
spectropolarimetric information to support such an interpretation, since the
slit had just reached the right edge of the filament at the moment that it
started to display such a motion.

How do the observed down- and up-flows fit into this scenario? The magnetic 
field lines in a flux rope twist about its axis by one or multiple windings. 
The filament plasma comes to rest in the dips produced by these winding field 
lines. Note that field lines closer to the axis of the flux rope produce 
only shallow dips, where the filament plasma is more susceptible to 
perturbations, while field lines lying further away from the axis have more 
pronounced dips, where the filament's plasma is maintained more stably. 
In Fig.~\ref{fig:model2} we show a flux rope indicated by two representative
field lines, one of them lying close to the flux rope's axis (green line), the
other further away (black line). They both form dips but possess different
degrees of twist. When the flux rope is in equilibrium (left panel), the
filament material, represented by the shadowed regions, rests inside the dips
of both field lines. Once the flux rope is destabilised and rises up (right
panel), the filament material still trapped inside the dips moves up with the
whole flux rope. This is observed as upflows which should be strongest near the 
centre of the rising flux rope according to the sketch. This material is 
trapped mainly in the outer, more twisted field lines. This rising material is 
associated with field that displays the opposite polarity to the underlying 
photospheric field. However, near the footpoints of the flux rope the
material may not be stably stored anymore, even in this outermost field line. 
It will flow down towards the legs. The magnetic polarity at such locations 
will be similar to that in the photosphere. Material bound to the less
twisted inner field line cannot be maintained stably against gravity in the
rising flux rope. It starts flowing down along the field line, contributing to
the fast downflows at both footpoints of the filament. These downflows are
associated with chromospheric field having the same polarity as the underlying
photospheric field. This qualitative model can explain the coexistence of up-
and downflows observed at the same spatial pixel positions as well as all the
inclination of the magnetic field vectors associated with these flows.

Field lines with different degrees of twist are present in every flux
rope (twisted flux tube). The twist increases from inside to outside, with a
completely untwisted field line in the centre, purely for symmetry
reasons. \citet{parker1,parker2} has described this structure and e.g., in the 
$2^{nd}$ order thin flux tube approximation \citep{pneuman,ferrizmas} it is 
clearly seen that the magnetic field azimuth, $\chi$, goes from zero at the 
axis to maximum at the edge. The green line in Fig.~\ref{fig:model2}
is an integral part of every twisted flux tube model of a filament or
prominence. Such a field line near the core of the flux rope can carry some of
the mass, although much of the prominence mass will lie below it \citep[unless 
the structure of the flux rope is similar to that described by][]{xu}.

Other scenarios are also conceivable that reproduce our observations. 
One of these is based on the results of by \citet{xu} and \citet{kuckein2}, 
who found plasma trapped both near the top and the bottom of the flux rope. 
Starting from a configuration similar to the one sketched in Fig. 10 of 
\citet{xu} in the quiescent state, the activation and rise of the filament 
would lead to the draining of material from the shallower dips, i.e. primarily 
from around the top of the flux rope (although material stored along the lower 
part of the more central field lines may also flow down the legs). Other 
scenarios are also conceivable. All conceivable scenarios have the 
disadvantage that they are more complicated than the one plotted in
Fig.~\ref{fig:model2} and this added complexity does not a priori improve the
reproduction of the observations. Therefore, following Occam’s razor, we
concentrate here on the simplest model, plotted in Fig.~\ref{fig:model2}.

This same behaviour, i.e. upflows near the centre of the rising flux
rope, where plasma remains trapped in dips, and downflows for field lines that
have lost their dips, has been found in recent simulations of the dynamics of 
cool plasma blobs in an erupting prominence \citep{suvanba}, although on a 
much larger scale. 

An interesting point is that in addition to the supersonically flowing gas, we
also find gas nearly at rest (or at least flowing only at subsonic speeds)
throughout the filament. At least in the inner part of the filament this gas is
associated with locally $\Omega$-shaped field lines. The presence of this gas 
may suggest that not all field lines expand and bend as the filament becomes 
unstable. Some parts (possibly lower-lying ones) may have maintained their 
geometry. 

Can we deduce the configuration from our observations? \citet{low3} proposed 
two different magnetic configurations of coronal flux ropes, one with the same 
sense of twist in the flux rope compared to the surrounding field (inverse 
configuration, see their Fig.~1), and another with the opposite sense of twist 
(normal configuration, their Fig.~2). Figure~\ref{fig:incsilicon} shows that 
at the photospheric level the magnetic field points upward (has positive
polarity) on the south-western side of the filament, and downward on the
north-eastern side, so that the magnetic field overlying the flux rope is
expected to go from SW to NE. However, the first two columns of
Figs.~\ref{fig:inc6030} show that the blue-shifted He components (the upward
moving material inside the flux rope) have positive polarity on the NE side of
the flux rope and negative polarity on the SW side. This is opposite from the
polarities of the surrounding field as indicated by the photospheric
measurements. This suggests that the observed filament has a sense of twist
that is opposite to that of the surrounding  arcade. Our observations are
consequently more consistent with the normal configuration. From the analysis
of the H$\alpha$ images we can say that the flux rope is sinistral
\citep[referring to the direction of the component of magnetic field along the
PIL as seen by an observer standing on the positive polarity side of the
filament, e.g.,][]{martin1}. To be consistent with our results and
conclusions, we show in Fig.~\ref{fig:model2} a sinistral left-helically 
twisted flux rope (the arrow on the black field line gives the sense of
twist) following the normal configuration of \citet{low3}. 

Most prominences have an inverse configuration, not a normal one 
\citep{bommier}. To our knowledge, the only evidence for flux ropes with 
normal configuration comes from Leroy's measurements of magnetic fields in 
prominences above the limb and very recently from the works of 
\citet{okamoto}, \citet{guo}, and \citet{kuckein1}. \citet{low3} claim that 
flux rope with normal configuration are responsible for fast CMEs from active 
regions. 
 
Note that in the interpretations given above the plasma having 
different LOS speeds within a single spatial resolution element is considered
to be lying above each other \citep[cf.][]{sasso1}. However, some of the plasma 
at different speeds may be at roughly the same height, e.g., if turbulence or 
strong counterflows are present in the plasma of the activated prominence.

\begin{acknowledgements}
We thank Aad Van Ballegooijgen for reading a draft of this paper and
generously providing suggestions. We thank the referee, Arturo L\'opez Ariste, 
for useful suggestions and comments. CS thanks the IMPRS on Physical Processes
in the Solar System and Beyond for the opportunity to carry out the research 
presented in this paper. This work was supported by the ASI-INAF contract ASI 
N. I/013/12/0, Work Package 1310 - Operazioni Scientifiche. This work has been 
partially supported by the WCU grant No. R31-10016 funded by the Korean 
Ministry of Education, Science and Technology.
\end{acknowledgements}

\bibliographystyle{aa}

\end{document}